\documentclass[twoside]{elsart} 
\usepackage{epsfig}
\usepackage{exscale}
\usepackage{amssymb}
\usepackage{subeqn}

\newcommand{\off}{\mathrm{off}}  
\newcommand{\out}{\mathrm{out}}  
  
\newcommand{\eff}{\mathrm{eff}}  
\newcommand{\dx} {{\mathrm d}x}  
\newcommand{\rad}{\mathrm{rad}}  

\newcommand{\sech}{{\mathrm{sech}}}

\newcommand{\Vc}{V_{0,c}}  
\newcommand{\II}{{I\! I}}  
\newcommand{\vII}{v_{\II,\out}}

\begin{document}
\begin{frontmatter}
\title{Nonlinear Schr\"odinger solitons scattering off an interface}
\author{Helge Frauenkron}
\address{\textit{HLRZ c/o Forschungszentrum J\"ulich GmbH\\
 D--52425 J\"ulich, Germany\\[0.5em]
 e-mail:  {\texttt{H.Frauenkron@fz-juelich.de}}}}

\begin{abstract}
  We integrate the one-dimensional nonlinear Schr\"odinger equation
  numerically for solitons moving in external potentials. In particular, we
  study the scattering off an interface separating two regions of constant
  potential modeled by a linear ramp. Transmission coefficients and
  inelasticities are computed as functions of the potential difference and
  the slope of the ramp. Our data show that the ramp's slope has a strong
  influence revealing unexpected windows of reflection in a transmission
  regime. The transmission coefficients for very small potential
  differences are compared with the theoretical predictions derived by
  perturbation theory.  Also the time evolution of the solitary waves after
  the scattering is studied. We observed that they in general behave like
  solitons with an amplified amplitude. Due to this, they oscillate. The
  oscillation period is measured and compared with theoretical predictions.

  PACS number(s): 03.40.Kf, 0.3.65.Ge, 42.65.Tg, 42.81.Dp
\end{abstract}

\begin{keyword}
Nonlinear Schr\"odinger equation, Solitons, External potential,
Transmission coefficients, Amplified solitons, Amlitude oscillation
\end{keyword}
\end{frontmatter}

\section{Introduction}
Recent years have seen a considerable growth in the interest for nonlinear
partial differential equations with soliton solutions. The most important
and very well studied equations are the Korteweg-de Vries equation with
serveral of its modifications, the sine-Gordon equation and the nonlinear
Schr\"odinger equation (NLSE). Historically, the first person who observed
a solitary wave was the scotsman J.S.\ Russell
\cite{russel:1838,russel:1844}. He discovered on the Edinburgh-Glasgow
channel that a large solitary `heap of water', as he calls this
phenomenon, moves over a very long distance without been dispersed. The
equation of motion to describe these waves in shallow water was first found
by D.J.\ Korteweg and G.\ de Vries in 1895 \cite{KdV:1895}. After this it
took some time, until 1967, when C.S.\ Gardener, J.M.\ Greene, M.D.\ 
Kruskal and R.M.\ Miura \cite{gardener:67,gardener:74} showed that the
Korteweg-de Vries equation can be solved analytically with the so-called
inverse scattering theory (IST).  Moreover, it was also possible to solve
the sine-Gordon equation and the NLSE with the aid of the same method
\cite{zakharov:72,ablowitz:73,ablowitz:74}.

In the present paper we study the NLSE which appears with its variants in
problems drawn from disciplines as diverse as optics, solid state, particle
and plasma physics. There, the NLSE describes phenomena such as
modulational instability of deep water waves \cite{ono}, propagation of
heat pulses in anharmonic crystals, helical motion of very thin vortex
filaments, models of protein dynamics \cite{fordy}, nonlinear modulation of
collisionless plasma waves \cite{lamb}, transmission of pulses through
junctions in optical fibers \cite{gordon,chbat,anderson:94}, and
self-trapping of light beams in optically nonlinear media
\cite{kivsh:90,kivsh:93}. In the last decade the optical communication via
solitons has become a field of special interest
\cite{hasegawa:73,mollenauer:80,fricke:92,kaertner:95}. In all these
applications, the main interest is due to the fact that the NLSE is the
most fundamental model which describes the interplay between weak dispersion
and nonlinearity for a wave envelope. The existence of soliton solutions is
one consequence of this. These are solitary waves with well defined
pulse-like shapes and remarkable stability properties \cite{jackiw}.

A great deal of current interest is directed to the question how solitons
behave under the influence of external perturbations
\cite{kivsh:90,karpman:79,bondeson:79,kivsh:87b,kivsh:89,nogami:94b,knapp:95}.
The perturbations describe e.g.\ various dissipative effects,
inhomogeneities of a medium or effects of external fields. They can be
naturally divided into two classes: Hamiltonian and dissipative.
Unfortunately, most of them destroy the integrability of the NLSE. But in
many cases, the perturbations are small and the techniques of analytical
perturbations theory can be applied to extract some results.
Perturbation-induced effects are of interest mainly because they represent
physical phenomena that cannot be comprised by exactly integrable models.

Here we shall limit ourselves to such perturbations which can be described
by potentials. They preserve the Hamiltonian structure of the NLSE
\cite{lamb}, but not its complete integrability. More precisely, we shall
consider the one-dimensional NLSE with external potentials which are
constant outside a finite interval. Thereby we emulate the effect of an
interface with finite width between two different media in which the
solitons have different characteristics. We study initial conditions
consisting of one single soliton which moves towards the interface. There
it will interact and, in general, it will not just be transmitted or
reflected. Instead, we expect that the soliton will also undergo inelastic
scatterings where it breaks up either into several solitons or into
nonsolitary waves, or both.

This problem has been studied previously by several authors. While
perturbative approaches were used in
\cite{kivsh:90,kivsh:93,kivsh:89,moura:94}, straightforward numerical
integrations were made in \cite{nogami:94b,nogami:94,knapp:95}. Both
approaches showed that the soliton behaves just like a Newtonian particle
if the force created by the potential is sufficiently weak. This is to be
expected, but the problem what happens when the force is strong was not
studied extensively. The problem with discontinuous potentials is
considered in \cite{giusto:84,minelli:85}, where the main interest was
broken conservation laws and soliton tunneling. In \cite{frauenkron:95} the
authors have studied the behavior of the soliton scattering at a potential
ramp only qualitatively. The main goal was the successful application of
higher order symplectic integrators to the NLSE. The spatial inhomogeneity
considered in \cite{kivsh:93} was more complicated, because the authors
allowed also a discontinuity in the nonlinear part. That lead to a
different behavior of the scattered solitons.

It is one purpose of the present paper to study the potential ramp
\cite{moura:94,chen:76} in a more quantitative and systematical way.  The
main reason to investigate this kind of external potential was the fact
that it reveals a very interesting feature. Namely, while for most
parameter values of the slope and the hight of the potential the soliton
behaves as expected in a ``particle-like" manner, for special sets of
parameters the soliton shows decidedly inelastic properties. In particular,
the nonlinear reflection and transmission coefficients of the mass and of
the energy are shown to possess windows of behavior that are quite
counter-intuitive at first sight. We find that as the width of the
potential ramp is varied, the transmission coefficient abruptly drops from
a finite non-zero value to near zero and then just as abruptly back to a
value near 1. This feature does not disappear if one smoothes the
potential. Only for very smooth potentials the behavior of the soliton
equals more and more that of a Newtonian particle in an equivalent
potential. First we have compared some of our numerical results for the
flat potential step with predictions by perturbation theory. They agree
very well in the expected validation range.

We also have extensively discussed the influence on the soliton parameters
(amplitude and velocity) during the scattering process. There we come to
the conclusion that the solitons escaping in both regions after the
scattering are ``amplified solitons". These are solitons whose amplitude is
multiplied by some constant factor. Their behavior was studied by Kath and
Smyth \cite{kath:95} who derived a formula to predict the amplitude
oscillation period depending on the soliton parameters. This formula is
applied to our data and is in quite good agreement with them.

This article is organized as follows. In the next section we briefly
summarize some basic relations and equations concerning the soliton
solutions of the NLSE and introduce our problem. A detailed discussion of
the transmission coefficients is made in Sec.~3. There some simulations are
compared with results from the perturbation analysis, as far as this is
possible within its validity range. In Sec.~4 we take a closer look at the
shape of the soliton after it is scattered at the inhomogeneity. An
amplified soliton ansatz is made and the parameters of the corresponding
solitary wave are determined. A summary and outlook is given in Sec.~5. In
the appendix the applied integration method (symplectic integrators) is
shortly described.  The advantage over the other explicit integration
schemes commonly used in this field, such as Runge-Kutta or
predictor-corrector methods, is explained.

\section{The NLSE soliton solution} 

Using appropriate units, we can write the NLSE as 
\begin{equation}
   i {\partial \Psi(x,t)\over \partial t} = -{1\over 2} 
   {\partial^2\Psi(x,t)\over \partial x^2} \,- \; |\Psi(x,t)|^2 \Psi(x,t) 
   + V(x)\; \Psi(x,t) ,             \label{nlse}
\end{equation}
where $V(x)$ is the external potential. We shall use for the latter a
piecewise linear ansatz, with $V(x)\equiv 0$ for $x<0$, $V(x)\equiv V_0>0$
for $x>x_0\geq 0$, and linearly rising for $x$ between 0 and $x_0$,
\begin{equation}\label{pot}
    V(x) = \left\{ \begin{array} {r@{\quad :\quad}l}  
        0 & x<0 \\
        x V_0/x_0\; & 0\le x<x_0 \\ 
        V_0 & x\ge x_0 \end{array} \right.. 
\end{equation}
We call the negative $x$-axis region $I$, while region $\II$ is the region 
$x>0$ (where $V(x)>0)$. 

The NLSE with the perturbation (\ref{pot}) is not completely intergrable. But
for a constant potential $V_0$ it is, and the soliton solutions of
Eq.(\ref{nlse}) form a two-parameter manifold (apart from translations).
Taking as parameters the velocity $v$ and the amplitude $a$, these
solutions read \cite{drazin}
\begin{equation}
   \Psi(x,t)= {a\over \cosh[a(x-v t)]} e^{i\{vx+[(a^2-v^2)/2-V_0]t\}} \;.
                  \label{soliton}
\end{equation}
We denote the velocity of the incoming soliton as $v_0$. Using a suitable
rescaling of $x,t$ and $\Psi$, we can always choose its amplitude as
$a_0=1/2$, without loss of generality.

Among the infinitely many conserved quantities (for $V(x)=\mathrm{
  const}$!) \cite{kivsh:89}, the following three are of particular interest:
\begin{enumerate}

\item [(i)]the mass or ``number of quasi-particles"
\begin{equation} \label{eqnN}
   N = \int |\Psi|^2\; \dx \; ,
\end{equation}
which is conserved due to phase invariance;

\item [(ii)]the energy 
\begin{equation} \label{E}
   E = \int \left( {1\over 2}\left|\partial\Psi\over\partial x
    \right|^2 -{1\over 2} |\Psi|^4 + V(x)\; |\Psi|^2 \right)\;\dx\; ,
\end{equation}
which is conserved due to time invariance;

\item [(iii)]and the momentum 
\begin{equation} \label{P}
  P = {1\over 2 i} \int \left( \Psi^{\ast}
    {\partial \Psi \over\partial x} - \Psi  
    {\partial \Psi^{\ast}\over\partial x} \right)\; \dx \; ,
\end{equation}
which is conserved due to translation invariance.
\end{enumerate}
For the soliton given by Eq.(\ref{soliton}) this yields
\begin{subeqnarray}
  N &=& 2a \,, \slabel{N_class}\\ 
  P &=& vN \quad\mbox{and} \slabel{P_class}\\
  E &=& (v^2/2-a^2/6)N + \langle V \rangle N \,, \slabel{E_class}
\end{subeqnarray}
where the average over $V(x)$ 
is taken with weight $\propto|\Psi|^2$ as indicated by Eq.(\ref{E}).

%It is easily seen that $N$ and $E$ are also conserved for nonconstant
%potential $V$, while this is not true for $P$. Denoting by $N_i$ $(i=I,II)$
%the mass in region $i$, we have thus $N_{I,\out}+N_{II,\out}=
%N_{I,\inn}= 1$. Similarly, energy conservation gives
%$E_{I,\out}+E_{II,\out}= E_{I,\inn} = (v_0^2-1/12)/2$.

In the cases where the soliton retains its shape, the motion of the soliton
can be understood as that of a Newtonian particle moving in the presence of
an effective potential \cite{kaup:78}. This assumption is valid when the
force created by the external potential is sufficiently weak. In our case
this is true when the local disturbance is small, i.e. $V_0\ll K_0$ or when
$x_0 \gg V_0$ in our case. Then we can apply Ehrenfest's theorem to get the
effective potential describing the motion of the Newtonian particle with
mass $m=2a_0=1$ and initial velocity $v_0$.  Thus, in the case where the
potential is much smaller than the kinetic energy of the particle $(V_0\ll
K_0, K_0=v_0^2/2)$ we expect transmission with a final velocity $\vII =
\sqrt{v_0^2-2V_0}$.  This follows from the energy conservation law and the
assumption that the soliton remains entire and no radiation is emitted.
This also predicts that there is no transmission if $v_0<\sqrt{2V_0}$
(i.e., $K_0<V_0$) and $x_0\gg V_0$. But we will see that this is more or
less the exception in our simulations and the true behavior will lead to
unexpected windows in the transmission coefficients.

%How far this picture of a Newtonian particle is valid will become clear
%later.

To get some more insight into the problem, let us first look at the motion
of a Newtonian particle with a shape $|\Psi(x,0)|^2$ in a potential
(\ref{pot}). We define the centre of mass of the extended particle
\begin{equation}
  \label{centreofmass}
  z(t)= {1\over N}\int_{-\infty}^{\infty} \!\! \dx\; x |\Psi(x,t)|^2 
\end{equation}
and derive an effective potential according to which the centre of mass
is moving. Starting from the equation of motion we get:
\begin{eqnarray}
  \label{ehrenfest}
    {d^2z\over dt^2} &=& -{1\over N}\int_{-\infty}^{\infty} \!\! \dx \; {dV(x)\over
      \dx} |\Psi(x,t)|^2 \nonumber\\[2pt]
    &=& -{1\over 2a}\int_0^{x_0} \! \dx \; {V_0\over x_0}|\Psi(x,t)|^2 \,,
\end{eqnarray}
where we have chosen the potential from Eq.~(\ref{pot}). While this is
still exact for solitary waves, we now approximate the wave function by an
unperturbed soliton centered at $z(t)$. The shape of the soliton is then
given by
\begin{equation}\label{density}
  |\Psi(x,t)|^2 = {a^2\over \cosh^2[a(x-z(t))]} \,.
\end{equation}
We are aware that this assumption is valid only if the local derivative
of the potential is very small. 

From the Eqs.(\ref{ehrenfest}) and (\ref{density}) we can derive the
effective potential
\begin{equation}
  \label{veff}
    V_{\eff} (z)= {V_0 \over x_0} \left[\ln\left({\cosh(az)\over
    \cosh(a(z-x_0))}\right)+{x_0\over 2}\right]
\end{equation}
according to which the centre of mass of a soliton (\ref{soliton}) is
moving. In the limit $ax_0 \gg 1$ one sees that $V_{\eff}(z) \to V(z)$. In
the upper limit $x_0 \to 0$ in which the potential becomes a step function
the effective potential takes the form
\begin{equation}
  V_{\eff} (z)={V_0 \over 2} (\tanh(az)+1) \,,
\end{equation}
which agrees with the potential derived by Nogami and Toyama \cite{nogami:94}.

As one can easily see these functions are smooth sigmoids in contrast to
$V(x)$. Due to this it is possible that the soliton's centre of mass
penetrates into the region $\II$ (i.e., where $z > x_0$) although
$V_0>K_0$. 

Because of the fact that the effective potential has no local minima the
soliton cannot be at rest at any place (provided that $v_0\ne 0$). This
statement does not hold if one also allows a change in the nonlinear term
at the inhomogeneity \cite{kivsh:90}. Then it is possible that the soliton
is trapped in the vicinity of the inhomogeneity.

\section{Transmission and reflection coefficients} 

Within our simulations we have intensively studied the transmission
coefficients of the mass, the momentum, and the energy. We found that they
strongly depend on the geometry of the external potential. Especially the
slope of the ramp has a very strong influence on them, which is not
expected for the equivalent particle method \cite{kaup:78}. For the
integration we used a fourth order symplectic integration algorithm
\cite{frauenkron:95} which is described in the appendix.

We measured the mass $N_i$ and the energy $E_i$ in each region ($i = I,\II$)
separately. Notice that $N$ and $E$ are conserved even in the presence of a
nonconstant potential \cite{giusto:84}. Thus, they are of special
importance among the infinite numbers of conserved quantities in the case
where the potential is constant \cite{kivsh:89}. Accordingly, we can define
two sets of transmission and reflection coefficients. We call them $T_N,
R_N$ and $T_E, R_E$,
\begin{equation}
   T_N={N_{\II}\over N}\;,\qquad R_N={N_I\over N}=1-T_N
\end{equation}
and
\begin{equation} 
   T_E={E_{\II}\over E}\;,\qquad R_E={E_I\over E}=1-T_E\; . 
\end{equation} 
A physical interpretation of $T_N$ can be the following: if $N$ is the
number of quasi-particles of the incoming soliton, then is $T_N$ the
quantity of quasi-particles penetrating the potential; resp. is $T_E$ the
energy of these quasi-particles.

Another quantity of interest is the momentum. It is only conserved for a
constant potential; otherwise the translation invariance is broken.
Nevertheless the momentum of the transmitted and reflected wave can be
defined in the same way
\begin{equation} 
   T_P={P_{\II}\over P}\;,\qquad R_P={P_I\over P}\; . 
\end{equation} 
Due to nonconservation, in general we have $T_P+R_P \neq 1$.

All these coefficients are functions depending on the incident velocity
$v_0$ of the soliton, the height $V_0$ and the length $x_0$ of the
potential ramp. Our main interest lays in the detailed dependence on these
parameters. Thus, we numerically evaluate the coefficients for a few
initial velocities of the soliton varying the length and the height over a
wide range.

Primarily, we will study these coefficients in a parameter range where the
perturbation approach is not valid. Only very few works in this field have
dealt with perturbations which are of the same order as the energy of the
soliton. Up to now the only way to get some insight is to simulate the NLSE
numerically.

In the following simulations we used $\Delta t=0.005 , \Delta x={\pi\over
  16}\approx 0.2$, and $a_0=1/2$. Hence the initial soliton takes
the form
\begin{equation} \label{init_soliton}
  \Psi(x,t=0)=\left\{2\cosh[{\mbox{$1\over
      2$}}(x-x_{\mathrm{off}})]\right\}^{-1} e^{i\, v_0\, x}\,,
\end{equation}
where the offset $x_{\mathrm{off}}$ is the position of the soliton at
$t=0$. The kinetic energy of the soliton is, according to Eq.~(\ref{E}),
$K_0= v_0^2/2$.

In addition to the transmission and reflection coefficients we also
registered all local maxima of $|\Psi(x)|^2$ with $|\Psi(x)|^2>1/3000$. The
threshold was chosen in such a way that no essential information is lost,
but small radiation and the noise which is produced due to the numerical
integration are not taken into account.

\subsection{Transmitted mass and energy}

\subsubsection{Potential step}

Let us first examine the potential step $(x_0=0)$, because there it is
possible to compare the numerically measured transmission coefficients with
the results calculated by perturbation theory. These calculations are
valid only in two situations: when the potential step is very low in
comparison to the kinetic energy of the soliton, and when the potential is
much higher than the kinetic energy (reflection at a hard wall). 

In the first case the soliton is transmitted nearly without loss of
radiation. The determination of the transmission coefficient is based on
the IST technique using a first order Born approximation. I.e., the
velocity of the soliton is assumed to be constant during the scattering
process. This is equivalent to the inequality \cite{kivsh:87b}
\begin{equation}
  \label{inequ}
  K_0 \gg 8a_0^2 |V_0|\,.
%  |V_0|a_0 \ll {v_0^2\over 8}\,.
\end{equation}

In order to calculate analytically the transmission coefficient for a
single soliton of the perturbed NLSE ($V_0 \ll K_0$)
\begin{equation}\label{nlse_eps}
   i {\partial\Psi(x,t)\over \partial t} = -{1\over 2} 
   {\partial^2\Psi(x,t)\over \partial x^2} \,- \; |\Psi(x,t)|^2 \Psi(x,t) 
   + V_0 f(x)\; \Psi(x,t) \,,            
\end{equation}
we follow the derivation given in \cite{kivsh:87}. There the
authors study a soliton which is scattered at an impurity ($f$ is a Dirac
$\delta$-distribution), whereas in our case the function $f$ is
the step function
\begin{equation}
    \theta(x) = \left\{ \begin{array} {r@{\quad :\quad}l}  
        0 & x<0 \\
        1 & x\ge 0 \end{array} \right.. 
\end{equation}
A calculation with this potential analogous to that in \cite{kivsh:87}
gives a formula for the amount of radiation which is emitted in the
backward direction,
\begin{equation}
  \hat{R}_N = \int_{-\infty}^0\!\!\!{\mathrm d}\lambda\, n_{\rad}(\lambda) \,,
\end{equation} 
and the energy of that radiation,
\begin{equation}
  \hat{R}_E = \int_{-\infty}^0\!\!\!{\mathrm d}\lambda\, 4\lambda^2 \,
  n_{\rad}(\lambda) \,,
\end{equation} 

Here $n_{\rad}(\lambda)$ denotes the total density of linear waves emitted
by a soliton during the scattering,
\begin{eqnarray} 
  n_{\rad}(\lambda) &=& {\pi V_0^2\over v^4}\; \beta^2(\lambda)\;
  \sech^2\left[ {\pi\over av} \left(2\lambda^2+{a^2\over 2}-{v^2\over
    8}\right)\right]\\\nonumber
  &&\times\int^{\infty}_{-\infty}\!\!\dx\int^{\infty}_{-\infty}\!\!\dx'\,
  f(x)\;f(x')\;\exp[i\beta(\lambda)(x-x')]\,,
\end{eqnarray}
where
$$  \beta(\lambda)={4\over v}\left[ \left(\lambda+{v\over 4}\right)^2 +
  {a^2\over 4} \right]\,.$$
The analytic calculation yields for the step function
\begin{equation}
  \hat{R}_N = {\pi V_0^2 \over v^4} \int_0^{\infty} \!\!{\mathrm
    d}\lambda \, \sech^2\left[ {\pi\over av} \left(2\lambda^2+{a^2\over
    2}-{v^2\over 8}\right)\right]\,,
\end{equation}
and 
\begin{equation}
  \hat{R}_E = {\pi V_0^2 \over v^4} \int_0^{\infty} \!\!{\mathrm
    d}\lambda \, 4\lambda^2 \,\sech^2\left[ {\pi\over av}
    \left(2\lambda^2+{a^2\over 2}-{v^2\over 8}\right)\right]\,.
\end{equation}
Inserting the parameters of the initial soliton ($v_0=0.8, a_0=0.5$) and
then calculating the integral numerically (using {\it Maple V}) it gives
the simple equations $\hat{R}_N \simeq 1.31 V_0^2$ and $\hat{R}_E \simeq
0.71 V_0^2$. This corresponds to a transmission coefficient for the mass
\begin{equation}
  \hat{T}_N =1-\hat{R}_N \simeq 1-1.31 V_0^2 \label{TNeps}
\end{equation}
and for the energy
\begin{equation}
  \hat{T}_E =1-\hat{R}_E \simeq 1-0.71 V_0^2\,. \label{TEeps}
\end{equation}

\begin{figure}[t]
  \begin{center}
    \epsfig{file=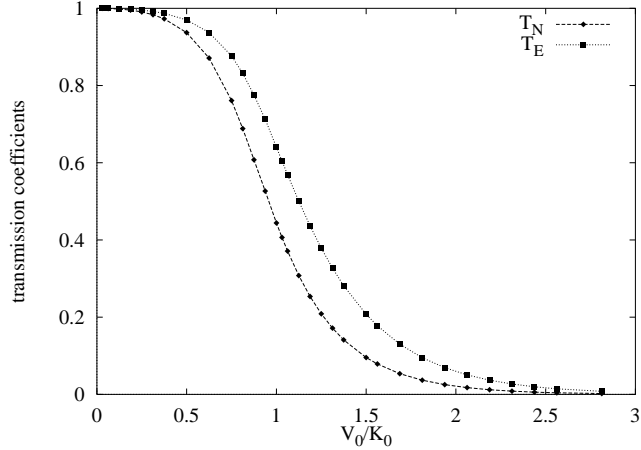,scale=0.7}
  \end{center}
  \begin{quotation}
  \caption{\footnotesize \label{trans0}\small The transmission coefficients
    $T_N$ and $T_E$ vs. $V_0/K_0$ with fixed $x_0=0$ and $v_0=0.8$. Errors
    are smaller than the symbols. The continues lines are drawn for guiding
    the eyes.}
\end{quotation}
\end{figure}

The quanitative behavior of the transmission coefficients measured from our
simulations is shown in Fig.~\ref{trans0}. The height of the simulated
potential step ranges from almost $0$ to about three times the kinetic
energy of the soliton. The transmission coefficients fill monotonously the
range from 1 to 0. We want to point out that the value of $T_E$ is always
larger than $T_N$.  This is due to the fact that the kinetic energy is
larger than the internal binding energy of the soliton ($3v_0>a_0$).

\begin{figure}[t]
\begin{center}
  \epsfig{file=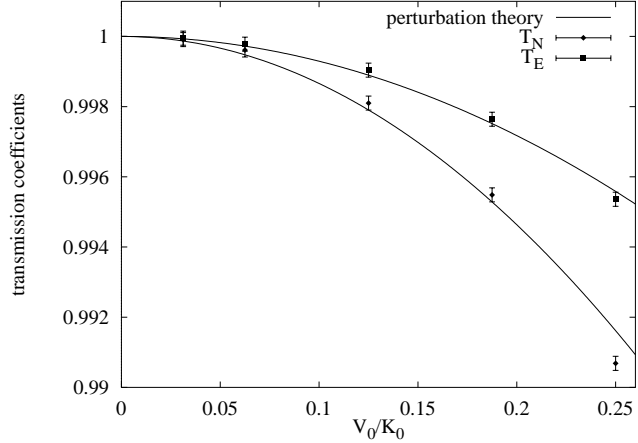,scale=0.7}
\end{center}
\begin{quotation}
\caption{\footnotesize \label{trans0_small}\small Comparison between the
  measured transmission coefficients $T_N$ and $T_E$ and the perturbation
  theory result $\hat{T}_N= 1-1.31 V_0^2$ and $\hat{T}_E= 1-0.71 V_0^2$,
  which is expected to be valid for $V_0/K_0 \ll 1/2$.}
\end{quotation}
\end{figure}
\begin{figure}[t]
  \begin{center}
    \epsfig{file=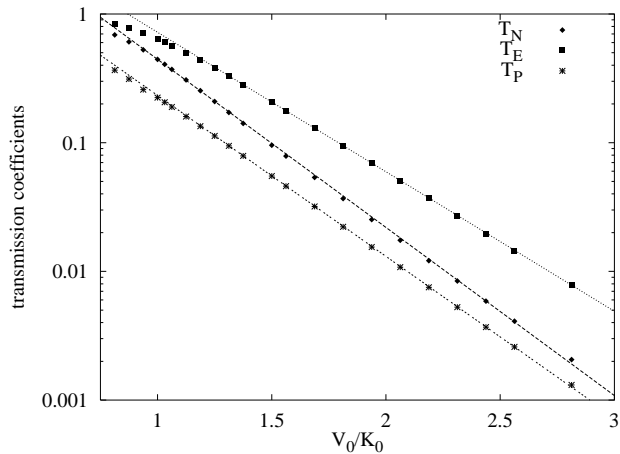,scale=0.7}
  \end{center}
  \begin{quotation}
  \caption{\footnotesize \label{trans0_exp}\small The transmission coefficients
    $T_N (\blacklozenge)$, $T_E (\blacksquare)$, and $T_P (*)$ vs. $V_0$ as
    in Fig.~\ref{trans0} in a logarithmic scale. An exponential function is
    fitted to each set of the data (points) to indicate its behavior for
    large values of $V_0/K_0$.}
\end{quotation}
\end{figure}

In Fig.~\ref{trans0_small} the result of perturbation theory (\ref{TNeps})
and (\ref{TEeps}) is compared with our numerically measured transmission
coefficient $T_N$ for small parameters $V_0$. Due to (\ref{inequ}) we
expect that the perturbation theory is valid in a range $V_0/K_0 \ll 1/2$
in the present case. In fact, one sees a rather good agreement between the
data and the first order perturbation theory for low step heights. The
discrepancy for larger perturbations ($V_0/K_0 > 0.25$) indicates that
higher order perturbation theory would be necessary to predict the
transmission of the soliton in this region more precisely. One has to bear in
mind that the perturbation analysis assumes that the initial soliton does
not break up and no soliton is reflected.

In the opposite case, where the soliton is reflected at very high potential
steps ($V_0\gg K_0$), we find an exponential decay of the transmission
coefficients (see Fig.~\ref{trans0_exp}). All three transmission
coefficients ($T_N$, $T_E$, and $T_P$) show this exponential decay for $V_0
> K_0$. The exponential functions plotted in Fig.~\ref{trans0_exp} are
fitted by hand. This simple behavior strongly suggests that one should be
able to derive it analytically. Up to now we did not yet succeed, but we
are still working on this problem. Intuitively it is clear that the higher
the potential is the less quasi-particles penetrate into region $\II$.
This amount of quasi-particles decreases exponentially with the penetrating
depth of the soliton, because the tail of the soliton decays exponentially.
Therefore the amount of quasi-particles ``pushed" into and staying in region
$\II$ decreases exponentially with increasing $V_0$.

\subsubsection{Potential ramp}
\label{pramp}

Let us now study potential ramps for positive values of $x_0$, i.e.
potential ramps with finite slopes. Our data show unambiguously that the
slope has a strong influence on the scattering process.  In the equivalent
particle method \cite{kaup:78}, where the soliton is treated like a
classical Newtonian particle with mass $N$ and kinetic energy $K_0$, one
would expect that the transmission coefficients remains constant when
changing the length of the ramp. This would be due to the constant
difference $V_0 - K_0$. But it turns out that this simple picture is not
true in our case.  Bearing in mind that the transmission coefficients for a
Gaussian wave packet of the linear Schr\"odinger equation being scattered
at a potential ramp shows a dependence of the slope, too, this is not that
remarkable. But the windows of reflection are.

\begin{figure}[b]
\begin{center}
  \epsfig{file=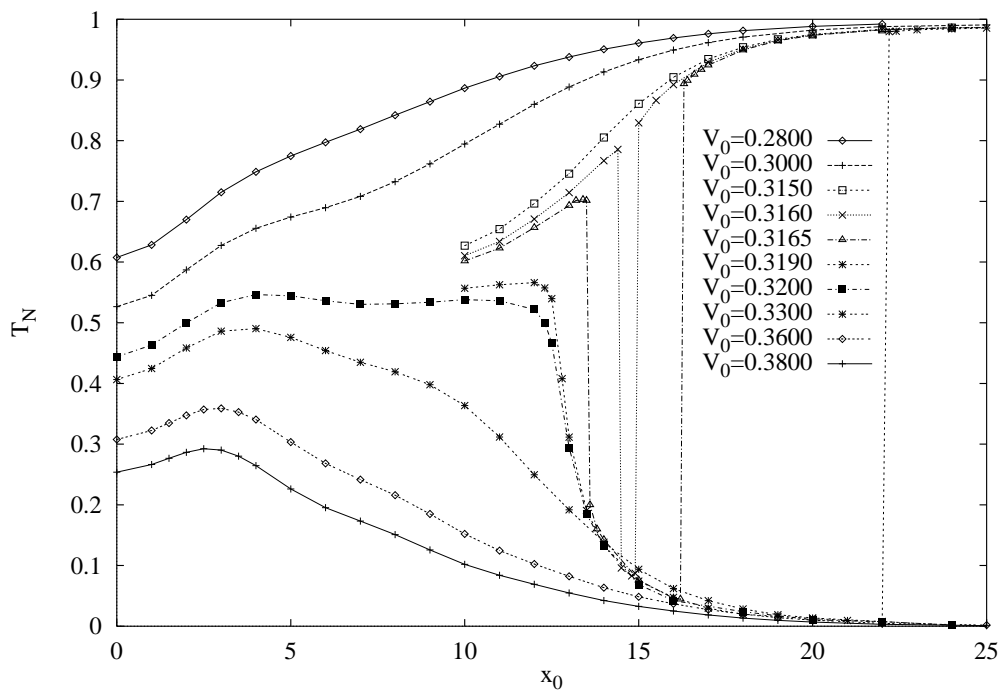,scale=0.85}
\end{center}
\begin{quotation}
\caption{\footnotesize \label{transNE}\small The transmission coefficient
  $T_N$ vs. $x_0$ for different potential heights and $v_0=0.8$.}
\end{quotation}
\end{figure}

In figure \ref{transNE} the transmission coefficient $T_N$ are plotted for
different slopes and heights of the ramp. A similar figure with less
complete data was already shown in Ref.\cite{frauenkron:96:1}. In the
following we will show mainly data for a initial velocity $v_0=0.8$ without
loss of generality. We also have simulated different initial velocities
$v_0$ and have found very similar results.

Let us first shortly recapitulate the observations found in
Ref.~\cite{frauenkron:96:1} concerning the global behavior of the
transmission coefficient. After that we want to discuss the windows of
reflection in a more detailed manner.  

The obvious stucture one observes is that the curves for $V_0 > K_0 = 0.32$
tend to 0 if $x_0\to\infty$, while those for $V_0 < K_0$ tend to 1 in
this limit. This indicates that the equivalent particle method is valid in
the limit of very flat ramps. If the kinetic energy of the soliton is
smaller than the potential height, it will be totally reflected; otherwise
it will be totally transmitted. Simulations with still flatter
ramps, e.g.~$x_0=100$ (not shown here), show that the transmission
coefficients are equal to $1$ resp.~$0$ within the error bars.

Another property is the nonmonotonic behavior of the transmission
coefficients at steep ramps. All curves $K_0\le V_0$ show a global maximum
at $x_0 \approx 3$.  If $V_0$ is decreased below $K_0$, the absolute
maximum is reached in the limit $x_0 \to\infty$ (that is 1), but there is
still a local maximum as long as $V_0\gtrsim 0.31$. At some threshold this local
maximum transforms into a shoulder. In Fig.~\ref{transNE} this threshold is at
$V_0\approx 0.31$ for $T_N$, while it happens at $V_0\approx 0.3$ for $T_E$
\cite{frauenkron:96:1}.  For all values of $V_0\lesssim 0.31$ the position of this
bump nearly stays at the same place.  Simulations with different initial
velocities ($v_0=0.2, 0.6$) likewise show such a maximum resp.\ shoulder at
$x_0\approx 3$. The fact that the maximum resp.\ the shoulder is located at
a length of the ramp which is roughly equal to the width of the soliton,
strongly suggests that this is a nonlinear resonance effect. A very similar
behavior of the reflection coefficient was found by Yu.S.\ Kivshar {\it et
  al.\/}\ \cite{kivsh:87} for solitons scattered by a single impurity.

Besides the global maximum, the curves for high potentials ($V_0\ge 0.36$)
show a slight shoulder at $5<x_0<10$. This shoulder gets more and more
pronounced if $V_0$ is decreased towards the magnitude of the kinetic
energy ($V_0=0.33$). At the same time its position is shifted towards
flatter slopes ($x_0\approx 7 \to 12$) and the shoulder becomes sharper.
In case the height reaches $K_0$ ($V_0=0.32$) this shoulder is a dominating
feature.

Decreasing the potential further (below $K_0$, but still above some
threshold $\Vc$), this shoulder continues to become sharper and more
pronounced. Indeed, it develops into a peak and the transmission
coefficients begin to grow with $x_0$ before they rapidly drop to join a
common curve. This curve seem to be independent of $V_0$ within the
interval $\Vc < V_0 < K_0$. But the transmission coefficients do not stay
on this common curve as we increase $x_0$ further. We observe that the
limit value for very flat slopes is not reached smoothly. Instead, the
transmission coefficients jump back{\bf ---}instantaneously within our
resolution{\bf ---}to values $\approx 1$.  The values of $x_0$, where these
jumps happen, seem to diverge when $V_0 \to K_0$ from below.

\begin{figure}[t]
\begin{center}
  \epsfig{file=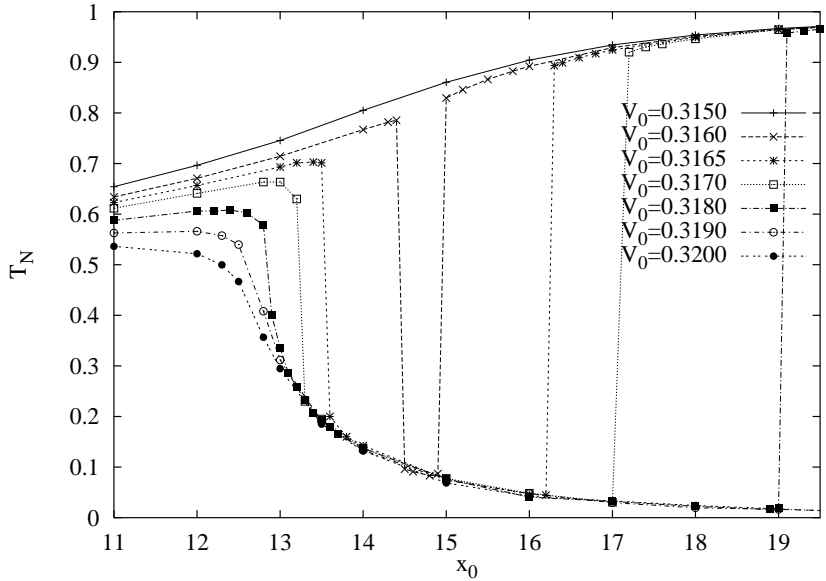,scale=0.9}
\end{center}
\begin{quotation}
\caption{\footnotesize \label{transNz}\small The transmission coefficient
  $T_N$ vs. $x_0$ as in Fig.~\ref{transNE}, but displaying additional curves revealing
  windows of reflection $(0.315 \le V_0 \le 0.32)$ with higher resolution.}
\end{quotation}
\end{figure}

As $V_0$ is decreased further, we reach a value $\Vc =0.3155\pm 0.0005$
where the window of small transmission coefficients disappears. As $\Vc$ is
reached from above, the locations of the downward jump and of the upward
jump both tend towards $x_{0,c} \approx 14.7$. At this point, both jumps
seem to be instrumentally sharp. For $V_0>\Vc$ only the second jump retains
this sharpness. For $V_0<\Vc$, no trace of this singular behavior is left,
and the transmission coefficients rise monotonically towards their
asymptotic value 1 as $x_0$ is increased. An enlargement of the relevant
part of Fig.~\ref{transNE}, together with some more data, is shown in
Fig.~\ref{transNz}. 

\begin{figure}[t]
  \begin{center}
%  \vglue-1.0truecm
%    \begin{minipage}[b]{8.0cm}
%      \epsfig{file=psi16.eps,scale=0.85}
      \epsfig{file=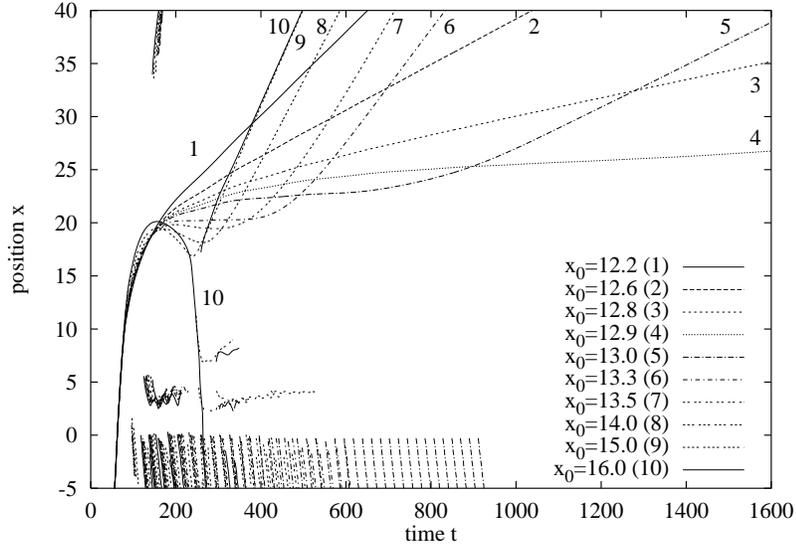,scale=0.85}
%    \end{minipage}
%    \hfill
%    \begin{minipage}[b]{4.8cm}
      \caption{\label{v318}      
        \footnotesize Time evolution of the local maxima of $|\Psi|^2$ for
        ten different slopes of the potential ranging from $x_0 = 12.2$ to
        $x_0=16.0$. The initial velocity is $v_0=0.8$ and the height of the
        potential is $V_0=0.3180 = 0.99375 K_0$. Only the maxima $|\Psi|^2
        > 1/3000$ are plotted, and only the most important maxima are
        labeled. The calculation was done on a lattice with $8192$ sites,
        discretization width $\Delta x\approx 0.2$ and integration step
        $\Delta t=0.005$.}\vspace{1.0truecm}
%    \end{minipage}
  \end{center}
\end{figure}

These unexpected and surprising windows of small transmission coefficients
were first seen in \cite{frauenkron:96:1}. This effect is not
expected within the picture of the equivalent particle method or within the
linear Schr\"odinger equation. Superficially similar effects have been observed within
the SG Eq.\ \cite{kivsh:91} and the $\phi^4$-model \cite{fei:92}. But there
are some fundamental differences to the actual work. These authors have
studied the effect of a local and attractive impurity which is such that
the soliton can be captured, and the energy is stored in an internal
mode. This leads to sharp resonance frequencies. Another difference is
that the soliton is unbreakable in \cite{kivsh:91,fei:92}. Thus, in
contrast to our system, the soliton is either reflected or transmitted or
captured (in a state oscillating around the impurity) and does not emit
radiation. The authors of \cite{kivsh:91,fei:92} observe that the soliton
may be reflected by the attractive impurity when its velocity lies in one
of several windows.  Within our resolution we find only \textit{one}
window, where the soliton is reflected instead of being transmitted. There
are no indications for further windows of reflection.

To understand this phenomenon in more detail, we first try to get some more
insight by looking at the evolution of the wave function. In
Fig.~\ref{v318} we show the local maxima of $|\Psi|^2$ for $V_0 = 0.3180 =
0.99375 K_0$. In this diagram one can directly compare the effect of
different slopes on the movement of the soliton. Let us begin with the
steep ramp $x_0 =12.2$ (label 1 in the figure). Here the soliton lowers its
speed while it climbs the potential ramp. At the same time radiation is
emitted in the forward and backward direction. The biggest amount of
radiation is emitted at the lower part of the ramp and immediately forms a
new soliton escaping in the backward direction. The slowing down process
continues until $t \approx 350$ and is accompanied with loss of radiation
in the backward direction.  After that the soliton moves with constant
velocity in the forward direction. But its mass has shrunk to roughly 60\%
of its initial mass.

Flattening the slope of the ramp, the velocity of the transmitted soliton
reduces continuously. Whereas its mass decreases only slightly. If the
slope is $x_0=12.8$ the soliton interacts with the ramp until $t\approx
1000$. This corresponds to a distance $x\approx 18$ between the maximum of
the soliton and the upper edge of the ramp. The same holds for the steeper
slopes described above.

Flattening the slope a tiny bit more ($x_0=12.9$) the soliton nearly
stops after penetrating in to region $\II$. It takes very long ($t\approx
1400$) until the soliton decides in what direction to move on. One
consequence is that more radiation is emitted into region $I$ while the
soliton interacts with the inhomogeneity. The mass shrinks to roughly 40\%,
and it takes until $t \approx 2700$ that the soliton accelerates to
full speed (not shown in Fig.~\ref{v318}).

\begin{figure}[t]
  \begin{center}
    \epsfig{file=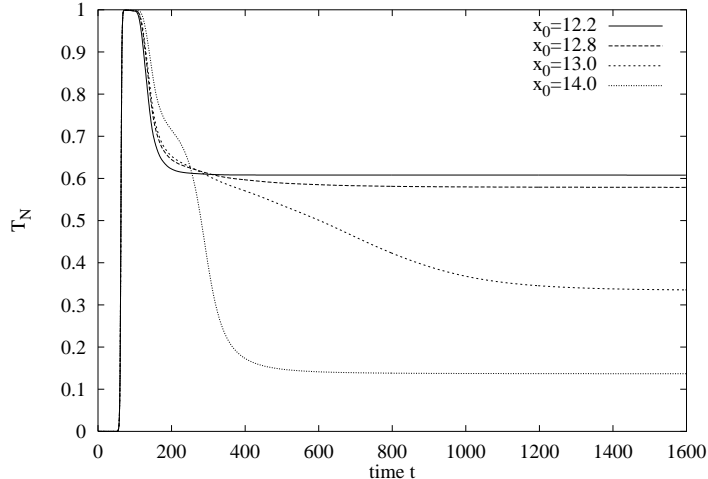,scale=0.75}
%    \hfill
%    \begin{minipage}[t]{4.8cm}
      \caption{\label{time_tn}\footnotesize Time evolution of the transmission
        coefficient $T_N$ of four representative values of $x_0$ (12.2, 12.8, 13.0
        and 14.0). The height of the potential is $V_0=0.3180 = 0.99375
        K_0$ as in Fig.~\ref{v318}}
%    \end{minipage}
  \end{center}\vspace{1.0truecm}
\end{figure}

This behavior changes if the slope is again only slightly flatter. For
$x_0=13.0$ the soliton moves with constant positive velocity already at
$t\approx 1100$. But its loss of mass due to radiation is much higher. The
transmitted soliton has only roughly 33\% of its initial mass. This loss
creates \textit{more than one} new soliton and nonsolitary waves escaping
from the ramp in the backward direction. It is hard to distinguish between
solitons and nonsolitary waves here, because the amplitudes are small and
furthermore the amplitude oscillates with a very low frequency, which makes
it hard to identify them as solitons.\footnote{The reader is refered to
  section \ref{sec:as} where the amplified solitons are discussed in more
  detail.} One should notice, that nonsolitary waves are created near the
lower edge of the ramp at least until $t \approx 900$. The other waves are
not plotted, because their amplitude is less than $1/3000$.  As one can see
from Fig.~\ref{time_tn} the transmitted soliton continuously loses mass
while it interacts with the potential ramp. This happens until it leaves
the upper edge of the ramp and moves with constant speed again.

If one looks closer at the traces of the transmitted soliton in
Fig.~\ref{v318} one observes small bends at times $t\approx$ 200, 350, 500
and 650. It looks as if the soliton's maxima are oscillating around a true
trace. This frequency corresponds exactly to the oscillation period of the
amplitude of those solitons. More precisely this means that the apparent
bend in the trace occurs when the transmitted soliton is maximally sharp.
This can be explained in the following way: the soliton is stretched when
it climbs up the ramp, which causes the amplitude to be reduced. Thereby
it becomes asymmetric as one can see in Fig.~2 of
Ref.~\cite{frauenkron:96:1} and its tail stays on the ramp for a long
time. During that time the pulsating soliton still interacts with the ramp,
which causes the oscillations seen in Fig.~\ref{v318}.

While decreasing the slope of the ramp more and more the mass of the
transmitted soliton resp.~solitary wave drops rapidly but continuously to
$\approx 10\%$ (see Fig.~\ref{time_tn}). This is the reason why the speed
of the soliton increases at the same time. For slopes $13.5 < x_0 < 15.0$
the movement of the penetrating soliton is not monotonical. After it has
come to a stop at $t\approx 200$ it moves back before it stops again and
escapes in the forward direction. E.g.~let us look at the situation with
$x_0=14.0$: at $t\approx 200$ the soliton becomes wider, which can be
concluded by comparing the oscillations of the other traces shown in
Fig.~\ref{v318}. Due to this broadening its maximum is shifted towards
region $I$. During this process some of its mass is pushed down the ramp
into region $I$. That can be concluded from Fig.~\ref{time_tn} where the
curve shows a drop at $T\approx 200$, after it had seemed to converge to a
value larger than that for $x_0=12.8$. After some part of the soliton has
separated from the bulk it accelarates away from the ramp while it
contracts again.

For $x_0=16.0$ the soliton penetrates into region $\II$, but is reflected
at the top edge of the ramp. This is consistent with the
collective-coordinate description (see Eq.(\ref{veff})). The reflected
soliton consists of roughly 96\% of the initial mass. This scenario holds
for flatter slopes, too, until $x_0 =19.1$. There the soliton is
transmitted with roughly 95\% of its initial mass and only few radiation
is emitted. 

For heights $V_0$ close to $\Vc$ the first jump becomes sharper and finally
as sharp as the second one.

As already pointed out in our previous work \cite{frauenkron:96:1} the
second upward jump can be explained easily: If there were no radiation (as
is the case for $x_0\to\infty$) the energy of the soliton would be
conserved, and the threshold for transmission would be strictly at $V_0 =
K_0$. But for finite $x_0$, the radiation implies that the soliton has lost
energy when it reaches the upper end of the ramp, and it needs $K_0>V_0$ to
be transmitted. Thus there is a range $K_0-\delta(x_0) < V_0 < K_0$ where
the soliton is reflected.  Since the soliton moves very slowly near the
upper egde of the ramp for $V_0\approx K_0-\delta(x_0)$, the dependence of
the transmission coefficients on the amount of radiation (and thus also on
$x_0$) is extremely sharp. To be more quantitative, let us consider $V_0 =
0.3190$. Here, the second jump (from nearly total reflection to nearly
total transmission) occurs at $x_0 = 22.1\equiv x_c$.  At this slope, $T_E$
jumps from 0.0093 to 0.9837, while $T_N$ jumps from 0.0074 to 0.9797.  The
fact that the soliton is coming to a near stop for $x_0=x_c+\epsilon$ means
that all its energy is potential energy, $E' = N' (V_0-a'^2/6)$, and the
energy difference has gone into radiation. Primes denote here observables
at the moment when the soliton stops. If we neglect the forward radiation
(which seems to be indeed small according to Figs.~\ref{v318}), we have $N'
= T_N$, $E' = T_E E_0$ and $a'=(T_N/2)^2$, and thus $T_E = T_N
(V_0-T_N^2/24) /E_0 = 0.9821$ for $x_0=x_c$. Comparing this with the
measured value 0.9837 we see good agreement.

Within the window of reflection, the time it takes until the point of
return is reached is not the same for different slopes. This time becomes
larger and larger the nearer the slope is to the border of the window.
Indeed, that is true for both borders. The reason for this is that the
radiative loss of energy is equal to the difference $K_0-V_0$.
Therefore the velocity of the temporarily transmitted soliton is very close
to 0. Thus, the soliton stays during a long time in the vicinity of the
egde of the ramp and can interact with it. This means that already tiny
amounts of radiation can cause the soliton to turn back into region $I$.
The same argument holds when decreasing the difference $K_0-V_0$. The
smaller it is, the longer it takes until the soliton has decided whether to
pass the potential ramp, or to fall back into region $I$.

In order to exclude that the observed feature is due to the
non-differentiability of the chosen potential, we have simulated smoother
potentials, too. We found that the values of the transmission
coefficients change only slightly if one uses instead of potential
(\ref{pot}) a more smooth sigmoid. Even the shape of the windows with small
transmission are still seen and they stay at the same positions. However,
if the potential becomes too smooth (e.g. such as functions like $\tanh(s
x)$ with $s<1$), the soliton behaves as one expects for the equivalent
particle method and only very little radiation is emitted.

\subsection{Transmitted momentum}

Another observable we can look at are the momenta of the reflected and
transmitted wave. Although momentum is not conserved for nonconstant
potentials, it is worth to have a closer look at it. If the wave function
is only composed of a single soliton, its momentum is directly related to
its velocity via Eq.(\ref{P_class}). In order to measure the momentum of
one soliton within a system of several nonoverlapping solitons, it is
necessary to restict the integration interval (in Eq.(\ref{P})) around its
position.  If we neglect the momentum of the radiation, and if at most a
single soliton is emitted in the forward and backward direction, it is
sufficient to partition the $x$-axis into regions $I$ and $\II$. That is
what we have done.

\begin{figure}[b!]
  \begin{center}
    \epsfig{file=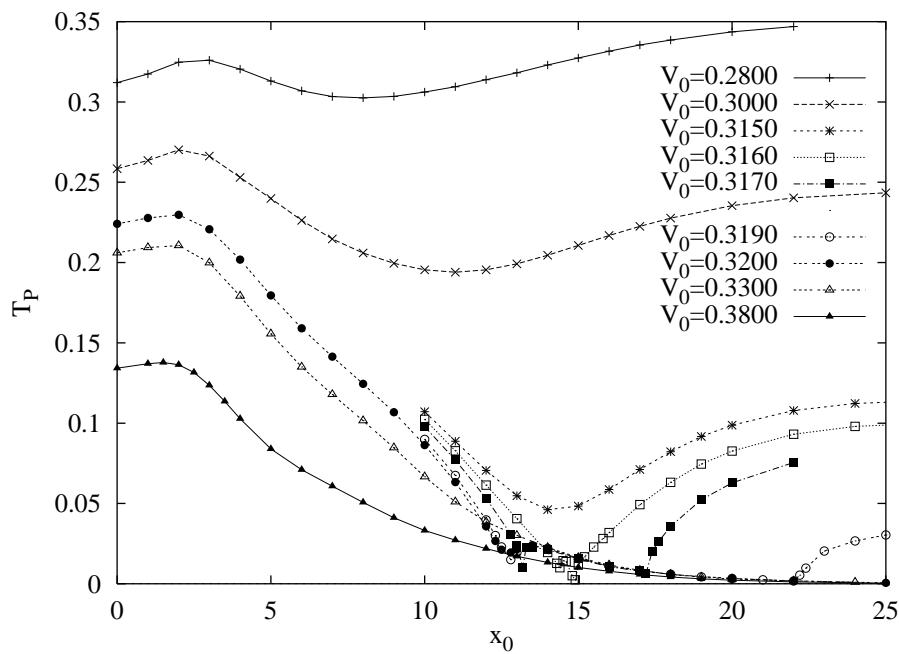,scale=1.0}
  \end{center}
  \begin{quotation}
  \caption{\footnotesize \label{transP}\small The transmission coefficient
    $T_P$ vs. $x_0$ for different potential heights as in
    Fig.~\ref{transNE}.}
\end{quotation}
\end{figure}

As mentioned before, according to the equivalent particle method the
transmitted momentum should not depend on the slope of the ramp. But what
we observe is different, as already pointed out in the previous section.
The global behavior of the transmission coefficient is the same as it is
for the transmitted mass and energy. In Fig.~\ref{transP} we see a (local)
maximum at steeper ramps ($x_0\approx 2 - 3$), which confirms the claim
that this is a nonlinear resonance effect. In the limit $x_0\to\infty$ the
momentum tends to the value one expects for the equivalent particle method.
There one assumes that if the soliton is transmitted, it is so without loss
of radiation. Due to the conservation of energy the reduced velocity is
given by $\vII$ in the cases where $V_0<K_0$. Otherwise the soliton is
reflected and the momentum of the transmitted solitary wave tends to
zero. Our results agree with this expected behavior for very flat ramps
($x_0 \gg 25$).
\begin{figure}[!tb]
  \begin{center}
    \epsfig{file=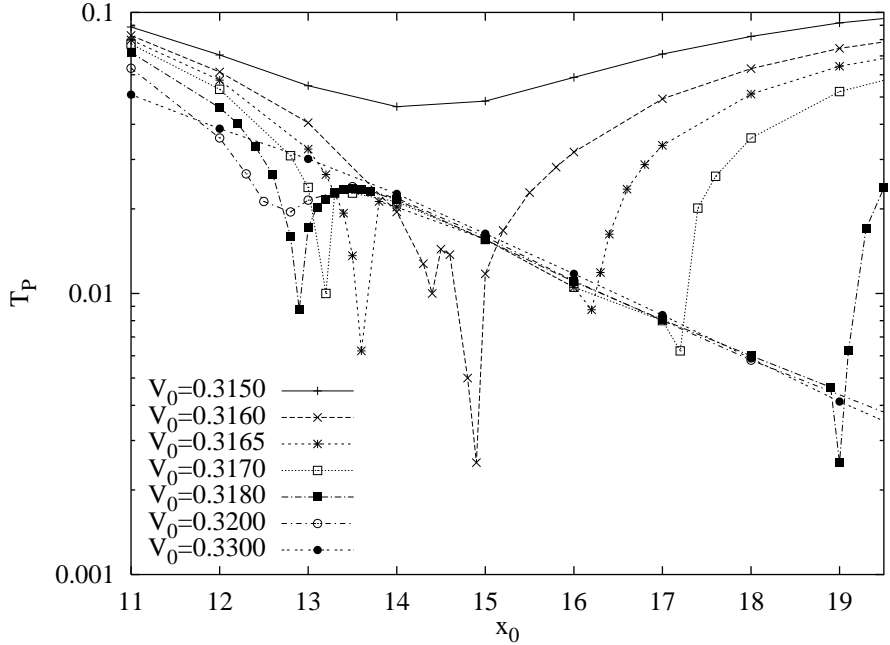,scale=1.0}
  \end{center}
  \begin{quotation}
  \caption{\footnotesize \label{transPmag}\small Enlargement of
    Fig.~\ref{transP} on a logarithmic scale. Only transmission
    coefficients for potential heights $0.3150\le V_0\le 0.33$ are
    plotted.}
\end{quotation}
\end{figure}

Between the (local) maximum and the limit region ($x_0\to\infty$) there
exists a minimum whose position depends on the height of the potential.
For lower heights it is located at steeper ramps ($x_0\approx 8$). While
raising the potential, the minimum gets more pronounced and is shifted to
flatter ramps ($x_0\approx 14$). When the ramp is nearly as high as the
kinetic energy of the soliton ($V_0\lesssim 0.3150$), the values of the
transmitted momentum at the minimum is very small and the minimum becomes
a dominating feature.

Raising the potential above the threshold $\Vc$ ($\approx 0.3155$; see
subsection \ref{pramp}) the situation is different.  For $V \in [\Vc,K_0]$
the momentum of the transmitted soliton has very sharp minima at both ends
of the window of reflection (see Fig.~\ref{transPmag}). In the interior of
the window it is non-zero and decreases exponentially with $x_0$. The
minimum at the upper end of the window tends to zero, and the same holds
for the lower end if $V_0$ is not too close to $K_0$. But for $V_0 \approx
K_0$ $(V_0\gtrsim 0.3185)$ when also the lower end of the window of
reflection in Fig.~\ref{transNz} is slightly smoothened, $T_P$ has a smooth
dip of finite depth.
%Flattening the slope of the ramp the momentum of the transmitted soliton
%smoothly tends to zero at the first edge of the window (see
%Fig.~\ref{transPmag}). Further on it jumps up to a finite value again. From
%there it decreases exponentially with raising $x_0$ until it reaches the
%second edge of the window. There it jumps to zero, before it smoothly---but
%with very large slope---increases again and converges to its limit value
%given by $\vII$. 
The larger $V_0$ is, the smoother the behavior gets.
Finally, it vanishes at a height $V_0> 0.325$. Even the curve for
$V_0=K_0=0.32$ shows a smooth dip at $x_0\approx 12.7$. The cusp at the
second edge remains sharp for values $0.3150 <V_0 < 0.3200$. Whereas the
first cusp shifts only little from $x_0=14.1$ to $x_0=12.7$ as $V_0 \to
K_0$, the second cusp shifts rapidly from $x_0=14.9$ to $x_0 > 22$. For
potentials higher than the kinetic energy we do not find any jump from
$T_P\approx 0$ to $T_P\approx 1$ at all.

Outside the window but still in its vicinity, the mass of the transmitted
soliton stays nearly constant. This leads to the conclusion that the
velocity of the transmitted soliton drops to almost zero at the edge of the
window. This agrees with the observation made in Fig.~\ref{v318}. The
dependence of $T_P$ inside the window is dominated by the mass of the
transmitted solitary wave, because its velocity changes only slightly. It
is conspicuous that all curves within the window match the exponentially
decaying curve for $V_0 =0.33$. This confirms the conclusion that those
transmitted solitary waves are universal in the sense that its mass, energy
and momentum are independent of the height of the potential.

%We have observed that the radiation in the backward direction is maximal
%at those values of $x_0$ where $T_P$ is minimal. The forward radiation does
%not show such a behavior. It is conspicuous that the transition is at the
%same values $V_0\approx 0.3155$ and $x_0\approx 14.6$ as it is seen in the
%scenario for $T_N$ and $T_E$.

\section{Shape of the reflected and transmitted soliton}

\subsection{Description by an amplified soliton ansatz}
\label{sec:as}
\begin{figure}[b]
%  \vglue -0.5cm
  \begin{center}
    \epsfig{file=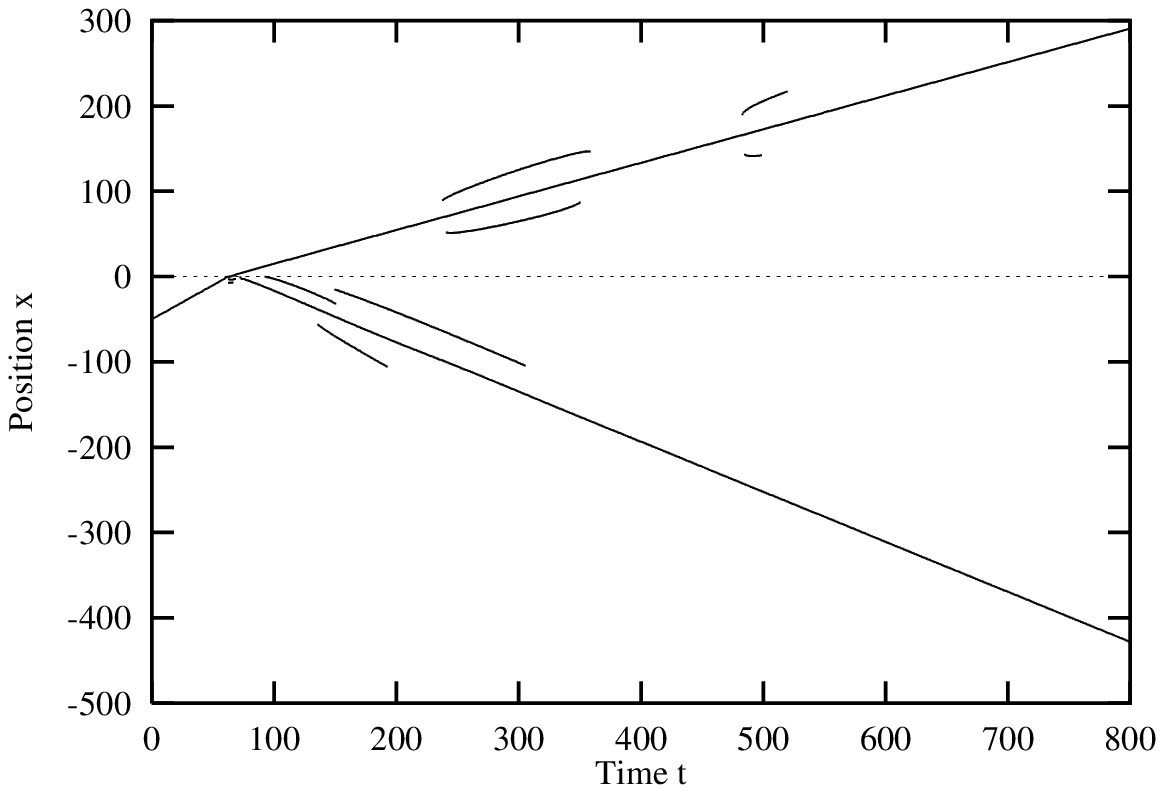,scale=0.75}
%    \hfill
%    \begin{minipage}[b]{4.8cm}
      \caption{\footnotesize \label{fig:trace30}Time evolution of local
        maxima of $|\Psi|^2$ for a soliton with incident velocity $v_0=0.8$
        which is scattered at a potential step with $x_0 = 0$ and height
        $V_0=0.3 = 0.937 K_0$. The calculation was done on a lattice with
        $4096$ sites, discretization width $\Delta x=0.2$ and integration
        step $\Delta t=0.005$. The latter parameters are the same for the
        next figures.}
%    \end{minipage}
  \end{center}
%  \vglue -0.5cm
%\end{figure}
%\begin{figure}[b]
%  \vglue -0.5cm
  \begin{center}
    \epsfig{file=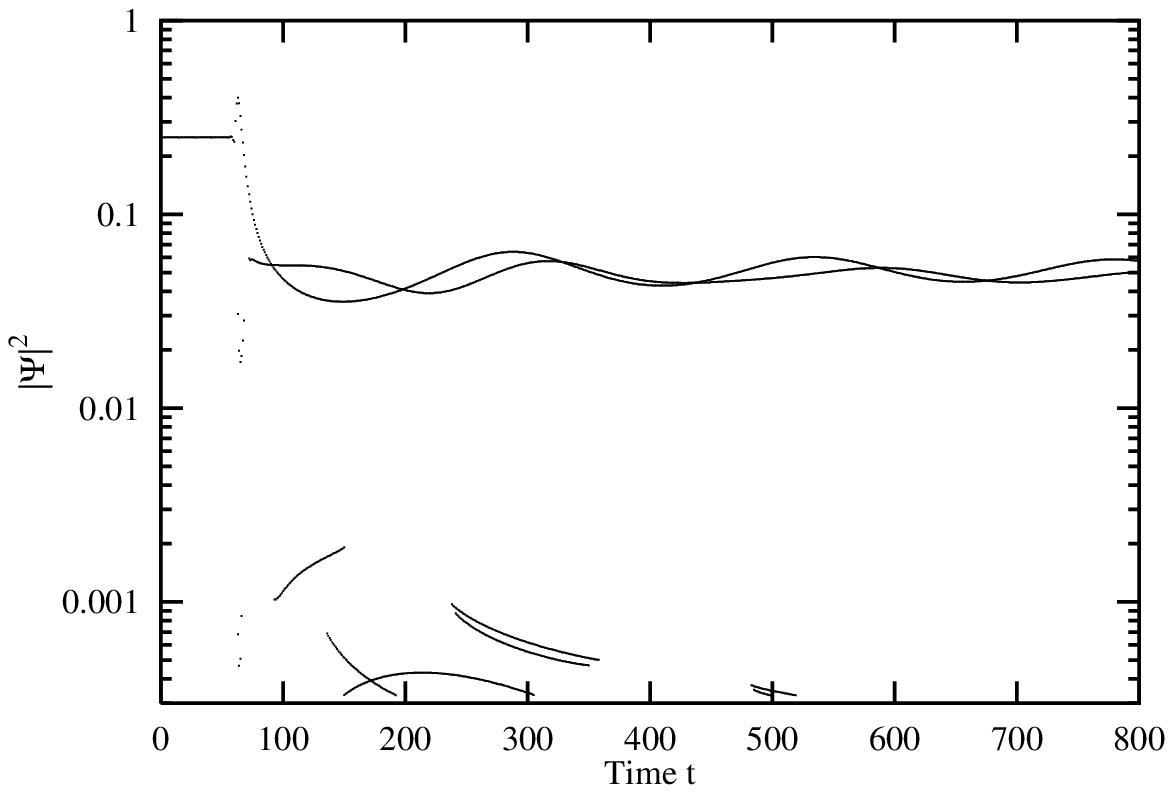,scale=0.75}
%    \hfill
%    \begin{minipage}[b]{4.8cm}
      \caption{\footnotesize \label{fig:height30}Time evolution of the
        height of the maxima shown in Fig.~\protect{\ref{fig:trace30}}. The
        highest curve belongs to the transmitted soliton, and the second
        highest to the reflected one. The other maxima presumably are due
        to the superposition of nonsolitary waves.\hfil}
%    \end{minipage}
  \end{center}
%  \vglue -0.5cm
\end{figure}

After the soliton hits the potential ramp, we observed in general that it
splits up and creates a transmitted and reflected soliton, which are both
accompanied by some radiation (see Fig.~\ref{fig:trace30}). This is why we
find typically more than a single maximum of $|\Psi(x)|$. Moreover, the
heights of these maxima in general are not constant in time, as one would
expect for a pure soliton state. Instead, the amplitude of the soliton
shows in general marked oscillations (e.g. Fig.~\ref{fig:height30}) which
are damped in all cases. In spite of this, the solitons moves with
practically constant velocity (see Fig.~\ref{fig:trace30}; the small bent
lines belong to the maxima of nonsolitary waves).

Damped amplitude oscillations have be observed in connection with
`amplified' solitons. These have been studied analytically and numerically
in \cite{satsuma:74,kuznetsov:95,kath:95}. These authors have shown that a
soliton whose amplitude is multiplied by some factor relaxes by emitting
radiation and by showing damped oscillations of its amplitude. This can be
understood by interpreting the initial state as a superposition of a
soliton with radiation.

As an example we just show one typical scenario to illustrate the movement
and the amplitude behavior of the solitons during the scattering process.
Other scenarios are shown in our previous work \cite{frauenkron:95}.
Figures \ref{fig:trace30} and \ref{fig:height30} depict the case where the
potential is a step function ($x_0=0$) and the kinetic energy ($K_0=0.32$)
is larger than its height $V_0=0.3$.  In the equivalent particle method one
would expect the soliton to move into region $\II$ and to propagate there
with a reduced speed $\vII$. But our simulation shows that it breaks up
into two solitons with roughly equal heights and with velocities $v=-0.588$
and $w=0.395$.  About half of the mass and three quarters of the energy are
transmitted ($T_N = 0.527, T_E = 0.712$).

The loss due to direct radiation is very small. If one inserts the above
numbers into Eq.(\ref{E_class}) one finds perfect agreement (discrepancies
are $\lesssim 1\%$) \cite{frauenkron:95}. This indicates that radiation in
form of nonsolitary waves is small in spite of the wiggles seen in
Fig.~\ref{fig:height30}.

Assume now that the wave functions after the interaction can be described
in both regions separately by amplified solitons
\begin{equation}
  \label{init:wave}
  \Psi(x',t'=0)= {a+\delta \over \cosh(a x')} e^{i\tilde{v}x'}\,,
\end{equation}
where $x'=x-\tilde{x}$ and $t'=t-\tilde{t}$. The parameters $\tilde{x},
\tilde{t}$ specify the location and the time of the interaction and
$\tilde{v}$ the new velocity. This function is a soliton whose amplitude is
simply multiplied by a factor of $\gamma = 1+{\delta\over a}$. The
long-time asymptotic solution consists just of a soliton. For finite times
it shows oscillations \cite{satsuma:74,kuznetsov:95}. They are damped, and
the soliton's amplitude tends to
\begin{equation}
  \label{fixed:ampl}
  \kappa = a+2\delta \; .
\end{equation}
The mass of the amplified soliton (\ref{init:wave}) at time $t'=0$ is given by
\begin{equation}
  \label{norm:0}
  N_0 := \int_{-\infty}^{\infty}\!\! \dx' \left|\Psi(x',t'=0)\right|^2 =
  2{(a+\delta)^2 \over a} 
\end{equation}
and its maximum is $|\Psi(0,0)|=a+\delta$. Satsuma and Yajima
\cite{satsuma:74} proved that the nonsoliton part of Eq.~(\ref{init:wave})
behaves like an ordinary wave packet and thus decays as $t^{-1/2}$ for
$t\to\infty$. The mass of the asymptotic soliton is given by
\begin{equation}
  \label{norm:infty}
  N_{\infty} := \int_{-\infty}^{\infty}\!\! \dx' \left|\Psi(x',t'=\infty)
  \right|^2 = 2(a+2\delta)  \,,
\end{equation} 
the difference with Eq.~(\ref{norm:0}) being due to the unobservable
radiation. 

Now we will use Eqs.~(\ref{norm:0}) and (\ref{norm:infty}) to evaluate
the parameters $a$ and $\delta$ for the scattered wave functions. Then we
will compare the scattering data with a sim\-ula\-tion using as initial
condition a wave function (\ref{init:wave}) with the calculated parameters.

The parameters $\tilde{x}, \tilde{t}, \tilde{v}$ have to be chosen in such
way that the amplified soliton fits to the scattering data. The velocity
$\tilde{v}$ of the scattered soliton can easily be extracted from a plot
that shows the time evolution of its maxima (see e.g. Fig.
\ref{fig:trace30}). However, the values of $\tilde{x}$
and $\tilde{t}$ are not fixed \textit{a priori\/}, because they depend on
the length of the ramp.

In the case of the potential step ($x_0=0$) it is natural to place the
interaction point simply at $\tilde{x}=0$, $\tilde{t}=-x_{\off}/v_0$, where
$x_{\off}<0$ is the position of the soliton at time $t=0$ (see
Eq.~(\ref{init_soliton})). But if $x_0>0$ the coordinates of the
interaction point are different from those.

To obtain the parameters $a$ and $\delta$ of the amplified soliton, the
equations
\begin{subeqnarray} 
  \label{solve:a:delta}
  N_0 &=& 2(a+2\delta+\delta^2/a)\\ 
  N_{\infty} &=& 2(a+2\delta) 
\end{subeqnarray}    
have to be solved. Here, $N_0$ is simply either the transmission or the
reflection coefficient, $T_N$ or $R_N$. The value of $N_{\infty}$ is taken
from plots which show the time evolution of the amplitude. From them it is
easy to estimate $\kappa=a+2\delta$, but the determination of $a$ and
$\delta$ themselves is less straightforward.

Solving Eqs.~(\ref{solve:a:delta}) we get two sets of solutions: 
\begin{subeqnarray}
  \label{solution:a:delta}
  a_{1,2} &=& N_0 - {N_{\infty}\over 2} \pm \sqrt{ N_0 (N_0-N_{\infty} )}\;
  ,\\ \delta_{1,2} &=& { N_{\infty} - N_0 \mp \sqrt{N_0 (N_0 - 
      N_{\infty})}\over 2}\; ,
\end{subeqnarray}
corresponding to initial amplitudes which are larger $(\delta_1<0)$ resp.\
smaller $(\delta_2>0)$ than that of the asymptotic soliton. Also, the
period of their oscillations are slightly different. 

\begin{figure}[t] 
  \begin{center}  
    \epsfig{file=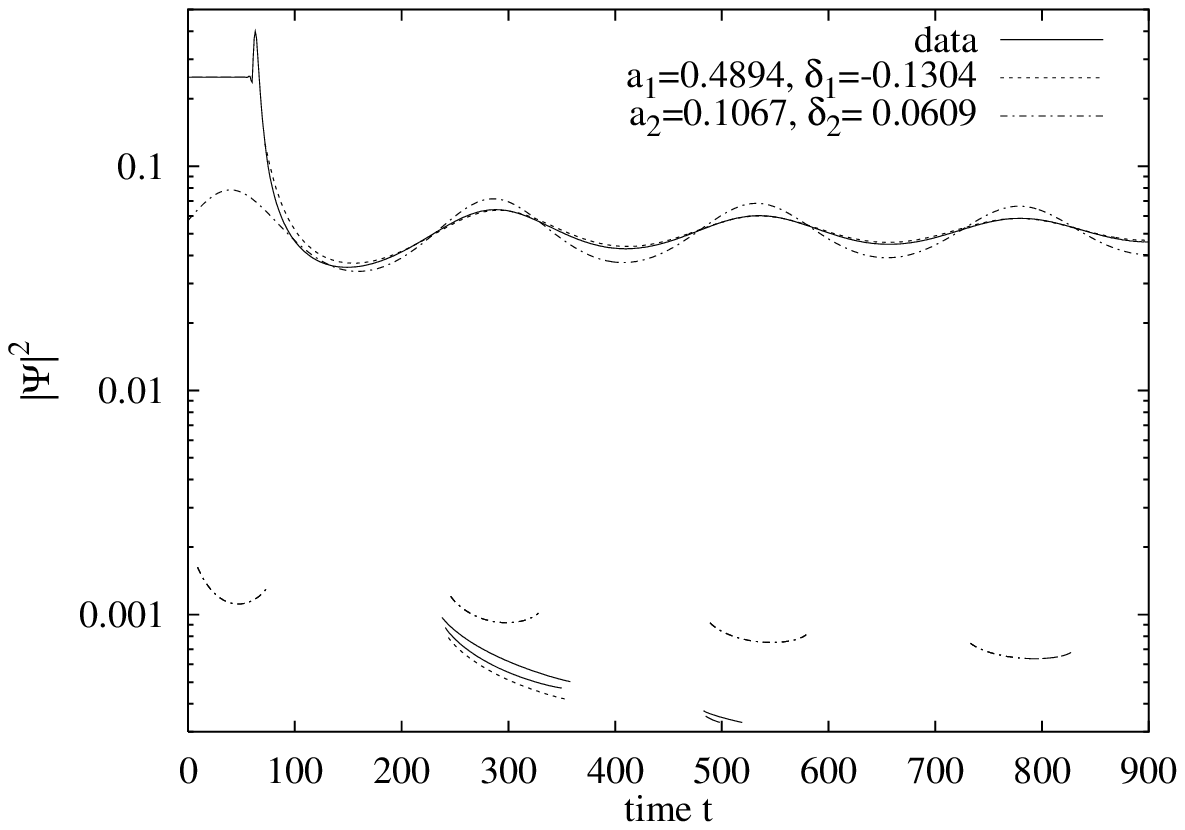,  scale=0.7}
  \end{center} 
  \begin{quotation}
  \caption{\footnotesize \label{fig:ad:x0_V3:ampl} Comparison between the
    amplitudes of the two simulated solutions $a_1=0.4894\pm 0.0400$,
    $\delta_1=-0.1304\pm 0.0200$, $a_2=0.1067\pm 0.0150$, $\delta_2=0.0609\pm
    0.0070$, and the transmitted soliton, Fig. \ref{fig:height30}) with the
    parameters $v_0=0.8 , x_0 = 0 , V_0=0.3$.}
\end{quotation}
\begin{center}
  \epsfig{file=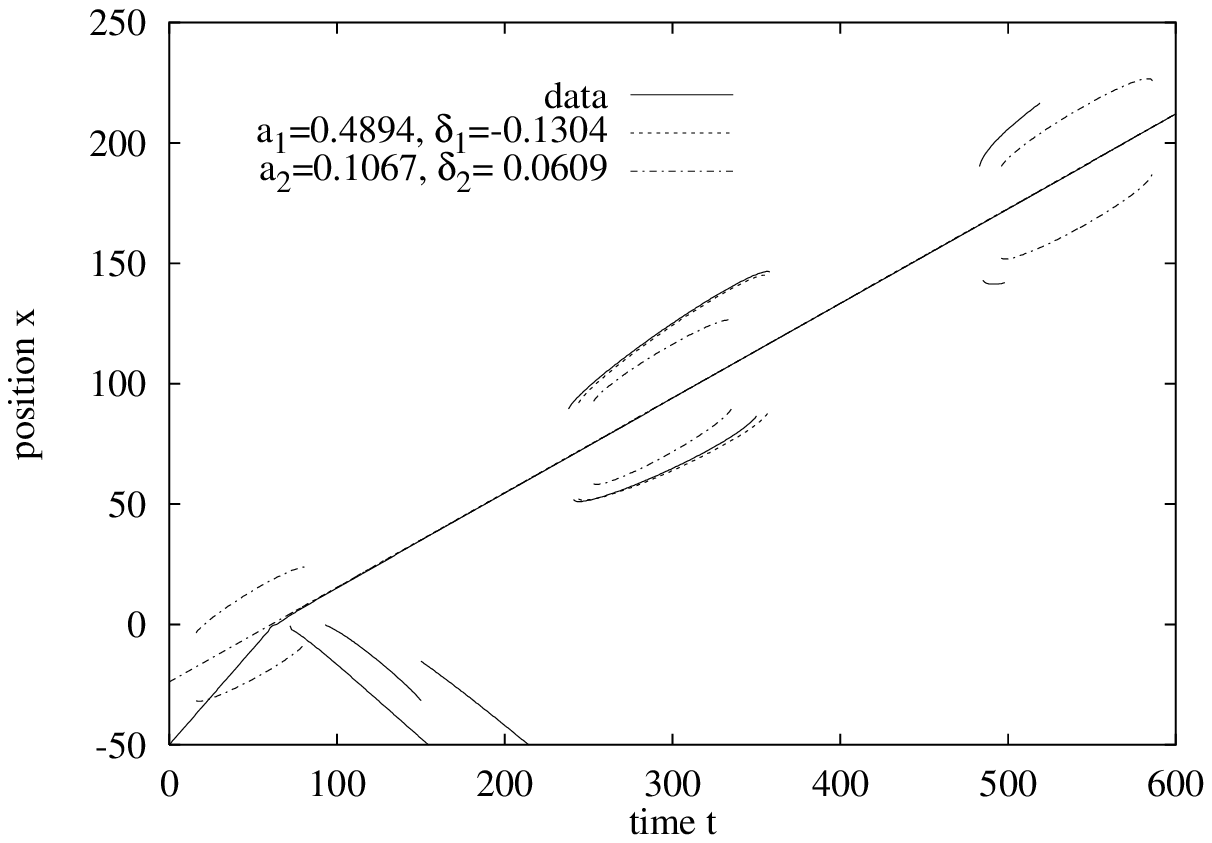,  scale=0.7} 
\end{center}
\begin{quotation}
\caption{\footnotesize \label{fig:ad:x0_V3:vel} Comparison between the
  position of the local maxima of the two simulated solutions and the
  transmitted soliton as in Fig.~\ref{fig:ad:x0_V3:ampl}.}
\end{quotation}
\end{figure}

Unfortunately, those relations do not help to decide which solution,
$(a_1,\delta_1)$ or $(a_2,\delta_2)$, describes the scattered soliton. This
can be found out by direct comparison, i.e.~numerically, because there does
not exist an analytical solution of the NLSE with initial condition
(\ref{init:wave}). As an example, we plotted the two solutions together
with the simulations in Figs.~\ref{fig:ad:x0_V3:ampl} and
\ref{fig:ad:x0_V3:vel}, for $v_0=0.8$ and $V_0=0.3$

As one can clearly see, only the amplified soliton with $a_1 = 0.4894 ,
\delta_1 = -0.1304$ is in good agreement with the data. Both amplified
solitons show oscillations with almost the same period as the transmitted
one. But the amplitude fits much better for the first solution, and the
behavior around the interaction point is much better represented. Also, the
side maxima are not that close to the main maximum as it is for the curve
with $a_2=0.1067 , \delta_2=0.0609$. All this is an stong indication that
the transmitted soliton can be described by an amplified soliton ansatz
with solution $a_1,\delta_1$.

Even better fits are obtained in Figs.~\ref{fig:ad:x0_V3:ampl} and
\ref{fig:ad:x0_V3:vel} if the parameter $a$ is not calculated from
Eqs.~(\ref{solution:a:delta}), but is simply chosen as the value
$a=a_0=0.5$ of the unperturbed soliton. Within errorbars, this is
consistent with the above value. The corresponding $\delta$ is $-0.13575$.

For this case ($v_0=0.8 , V_0=0.3 , x_0=0$) the interaction of the incoming
soliton with the potential can be interpreted in the following manner: the
soliton with amplitude $0.5$ interacts with the potential inhomogeneity
($x_0=0$). The main part $(T_E\approx 0.7)$ is transmitted, and the
remaining part is reflected. Most of the radiation is emitted in the
backward direction. Due to the birth of a second soliton the amplitude of
the initial soliton is reduced by $\delta$.  This new state of the wave
function is not a pure soliton any more. Thus, the amplitude decreases
further and begins to oscillate in the manner seen in the figures. The
reflected wave also forms such an amplified soliton. In addition, radiation
is created as seen from the various decreasing and non-oscillating side
maxima.

For flatter slopes of the ramp, amplified solitons fit well for times $t'$
larger than one oscillation period of the amplitude. The early stages of
the interaction can not be reproduced with this ansatz. In our simulations
we observed that the amplitude oscillation is slightly compressed
horizontally while the soliton is located in the rising part of the
potential.  This might be because of the transfer of energy into the
radiation.
\begin{figure}[t]
\begin{center}
  \epsfig{file=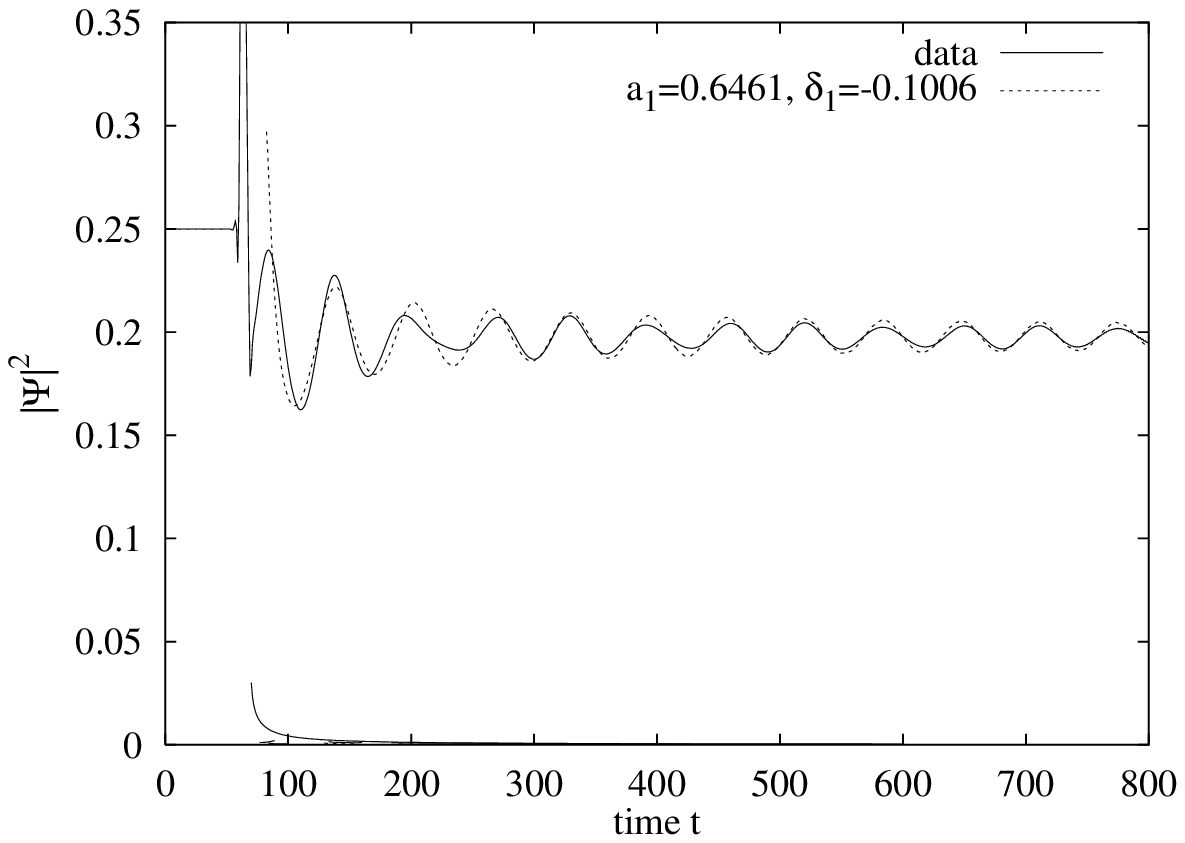,scale=0.7} 
\end{center}
\begin{quotation}
\caption{\footnotesize \label{fig:ad:x0_V5:ampl} $|\Psi|^2$ of the simulated
  amplified soliton with $a_1=0.6461\pm 0.0400$, $\delta_1=-0.1006\pm
  0.0100$ in comparison with the reflected soliton with
  the parameters $v_0=0.8 , x_0=0 , V_0=0.5$.}
\end{quotation}
                                %\end{figure}
%\begin{figure}[h!]
\begin{center}
  \epsfig{file=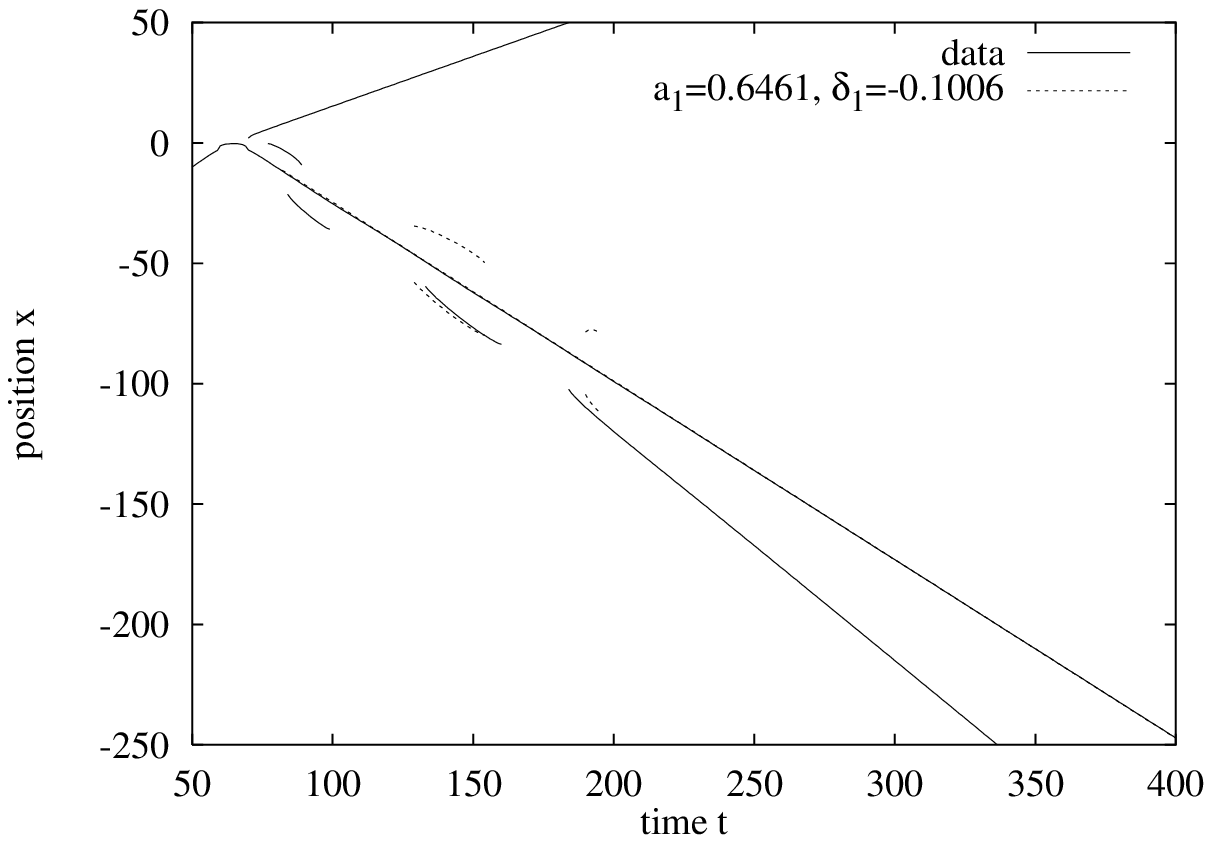,scale=0.7} 
\end{center}
\begin{quotation}
\caption{\footnotesize \label{fig:ad:x0_V5:vel} Comparison between the
  position of the local maxima of the simulated solution and the reflected
  soliton as in Fig.~\ref{fig:ad:x0_V5:ampl}.}
\end{quotation}
\end{figure}

The amplitude oscillations for the reflected soliton are less easy to
compare. This comparison is possible for very steep ramps and rather high
potentials. In those cases the reflected soliton has a larger amplitude
than the nonsolitary wave emitted during the interaction. Thus, effects due
to interferences are rather small. If their amplitudes are of the same
order, they strongly interfere and deform the shape of the amplified
soliton.  A way to avoid that problem is to wait very long until the
radiation either has passed through the soliton or is dispersed. But this
would require very long simulations. Another problem is that in general
the oscillation periods of the reflected soliton get very large which
requires a very long simulation time, too.

Another comparison between data and theory is shown in
Figs.~\ref{fig:ad:x0_V5:ampl} and \ref{fig:ad:x0_V5:vel} for the situation
$v_0=0.8$, $V_0=0.5$, and $x_0=0$. The simulation and the amplified soliton
show only small discrepancies, mainly during the first three oscillation
periods.  The interference is also the reason why the side-maxima of the
radiation emitted by the amplified soliton are not reproduced correctly.
The rest is quantitatively very well reproduced.

In summary, each of the solitary waves after the scattering can be
described as an amplified soliton, except for a very short time during the
collision process. It is described by means of only two parameters:
the reflected resp.\ transmitted mass and the height of the soliton in the
limit $t\to\infty$. Since these states are not solitons at finite times,
they emit radiation. This happens continuously and the amount radiated
until a time $t$ is proportional to $\sqrt{t^{-1}}$ \cite{satsuma:74}.
Indeed, we see that the amplitude of the linear waves does not decay
exponentially \cite{kuznetsov:95}. In the limit $t\to\infty$, the solitary
waves converge to solitons with defined shapes. The total amount of
radiation lost is simply
\begin{equation}
  N_0-N_{\infty} = 2{\delta^2 \over a} \,.
\end{equation}
But this means that the overall amount of radiation produced during the
whole scattering process is in general much bigger than one might have
supposed at first sight. For steeper ramps more than 10\% of the mass and
even slightly more of the energy is eventually radiated by the two
amplified solitons leaving the ramp into two directions. Summing up all
effects we find that in cases where the potential is of the same order as
the soliton's kinetic energy and the slope is steep, about one fourth of
the initial mass and energy is radiated. Otherwise the loss is smaller. The
percentage of radiation decreases systematically with decreasing slope.  In
the limit of very flat ramps $(x_0\to\infty)$ it tends to 0.

\subsection{Oscillation period of the solitons amplitude}

Kath and Smyth \cite{kath:95} derived two formulas which estimates the period
of the amplitude oscillations for an amplified soliton with
shape~(\ref{init:wave}). For this, they used an alternative method to
describe the time evolution of the soliton, namely a Lagrangian approach
\cite{anderson:83}. This leads to four approximate differential equations.
These coupled ODE's allow to predict the time evolution of the soliton's
parameters including the ``mass loss" due to dispersive radiation.  Their
results are in agreement with the full numerical solution of the NLSE.

As we have shown above, the assumption that the solitary wave after the
scattering is described by an amplified soliton with suitable values for
$a$ and $\delta$, seems to be valid in a wide range of external
perturbations. Thus we can apply their formulas to our data to see how far
they agree.

For small perturbations, it was shown in \cite{kath:95} that the period of
the amplitude oscillations is the same as that of the phase oscillations of
the pure soliton at time $t=\infty$:
\begin{equation}
  \label{Td0}
  T \simeq {4\pi \over \kappa^2}\;.
\end{equation}
For finite perturbations $(|\delta|>0)$, the same authors found (using our
notation)
\begin{equation}
  \label{period}
  T={\pi^2 \over \kappa^2} \left[ 1-\left( 1-{a^2 \over (a+\delta)^2}
\right)^2 \right]^{-{3\over 2}}.
\end{equation}
Notice that this does not agree with Eq.(\ref{Td0}) for $|\delta|\to 0$,
as the authors of \cite{kath:95} already pointed out.

\begin{figure}[tb]
  \begin{center}
%    \vglue -3mm
    \epsfig{file=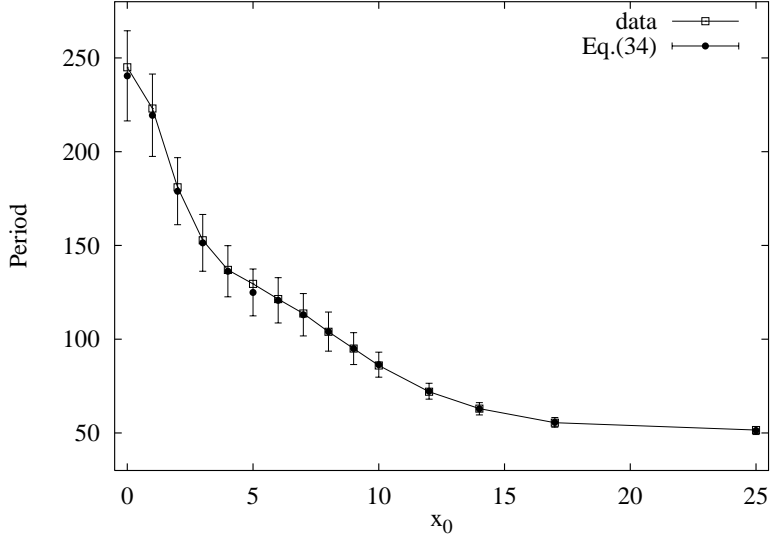,scale=0.85}
  \end{center}
%  \vglue -5mm
  \begin{quotation}
  \caption{\footnotesize \label{frqv8}\small Oscillation periods of the
    transmitted soliton (data), compared with the ones calculated using
    Eq.(\ref{Td0}). Small values of $x_0$ correspond to large $|\delta|$
    and large $x_0$ to small $|\delta|$. The initial parameters of the
    system are $V_0=0.30$, $v_0=0.8$ and $a_0=0.5$. The errors on the
    directly measured data are not shown, but are roughly half as big as those
    plotted for Eq.(\ref{Td0}).}
\end{quotation}
\end{figure}

\begin{figure}[tb]
  \begin{center}
%    \vglue -3mm
    \epsfig{file=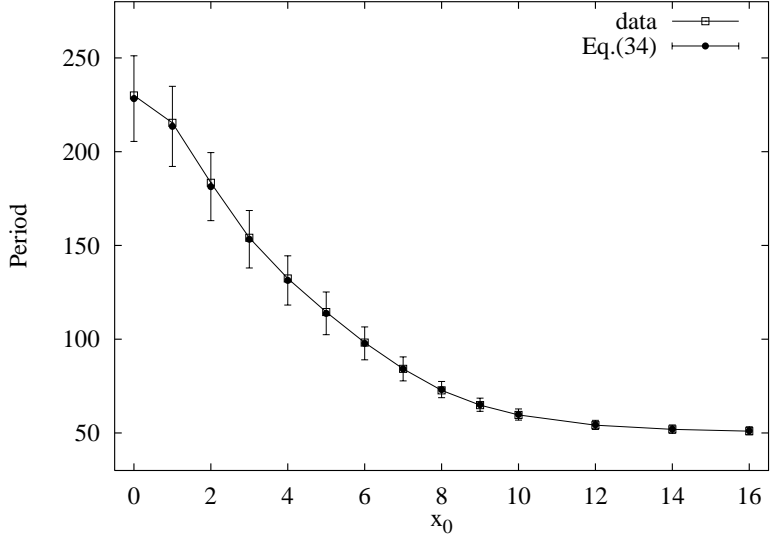,scale=0.85}
  \end{center}
%  \vglue -5mm
  \begin{quotation}
  \caption{\footnotesize \label{frqv6}\small The same as in
    Fig.~\ref{frqv8} but for different initial parameters ($V_0=0.17$,
    $v_0=0.6$, $a_0=0.5$).}
\end{quotation}
\end{figure}

Measured oscillation periods are plotted in Figs.~\ref{frqv8} and
\ref{frqv6} against $x_0$ for two different velocities. These are
comparisons with the predicted values of Eq.(\ref{Td0}), where we have
substituted $\kappa =a+2\delta$. The values predicted by Eq.(\ref{period})
do not fit the data in a sufficient manner. It predicts smaller values for
$x_0>16\,\, (v_0=0.8)$ resp.~$x_0>10\,\, (v_0=0.6)$ and much larger values
for $x_0\to 0$. But the simple Eq.(\ref{Td0}) shows a surprisingly good
agreement for all data points in both cases.

We could conclude from this that the disturbance of the soliton's amplitude
we are dealing with seems to be not as large as expected. Namely, in the
sense that Eq.(\ref{Td0})\textbf{---}meant to be valid only for very small
disturbances\textbf{---}predicts the correct time periods. This is
astonishing, because we found an reduction of the soliton's amplitude up
to 25\%! This would mean that the valitiy range of Eq.(\ref{Td0}) is much
larger than assumed in \cite{kath:95}.

%In both figures one sees that the data and the theory show qualitatively
%the same behavior. For small perturbations $|\delta|$ ($x_0\gg 1$), the
%meassured values of $T$ are larger than the values calculated by
%Eq.(\ref{period}), but agree perfectly with Eq.(\ref{period:4}). Going to
%larger perturbations the data coincide with Eq.(\ref{period}). For even
%larger perturbations $|\delta|$ ($x_0 \simeq 1$) even this formula gives
%too larg values.  But they still agree within the error bars, which also
%grow in this case. These discrepancies show that Eq.(\ref{period}) has a
%limited range of validity as one might have expected, and does not hold if
%the perturbation of the initial soliton is large.

%This qualitative and, except for large perturbations, quantitative
%agreement confirms two points: after the scattering the wave function
%consists of one or two amplified solitons, and the formula derived by Kath
%and Smyth is approximately valid for not too large perturbations.

\section{Summary and conclusions}

In this paper we found that solitons break up in general when hitting a
potential threshold. If the slope is very steep and the potential height is
either very low or very high compared to the soliton's kinetic energy
$(K_0\gg V_0 \mbox{ or } K_0\ll V_0)$, the soliton acts like a Newtonian
particle. The same behavior is seen if the slope is rather flat
$(V_0/x_0\ll 1)$ for most values of $V_0$. This includes the fact that in
those cases only very little energy and mass is transfered into radiation.
In such cases only one soliton with nearly constant height exists during
the whole simulation time. The amplitude of this solitons shows small
oscillations with a period, that is the period of the phase oscillation
of the undisturbed soliton.

The complexity of the outgoing state depends on the parameters of the
potential and of the soliton, but most frequently the soliton breaks into
two solitons of different mass. We have shown that these solitons can be
described by amplified solitons whose parameters $a$ and $\delta$ can be
evaluated from the data. Additionally, the period of the amplitude
oscillation can easily be calculated. The scattered solitons are
accompanied by radiation, which is emitted continuously and whose amplitude
decays proportional to $1/\sqrt{t}$ in time for large times. The fact that
the radiation decays only with a power law raises some problems with the
transmission of information via optical fibers \cite{kuznetsov:95}.

In some region of the ($x_0$, $V_0$)-plane the soliton is almost fully
reflected even if its kinetic energy is slightly larger than the potential.
This is due to the emission of radiation. The energy of the soliton is
reduced thereby, whence it cannot overcome the ramp and is reflected. The
associated jumps in the transmission coefficients are rather sharp and the
slope where this happens depends on the hight of the potential. The window
in slopes where the soliton is reflected shrinks if the height of the
potential is lowered.  At a critical value $\Vc$ this feature disappears
and the transmission coefficients rise monotonically. As the potential
height tends to the value of the soliton's kinetic energy, the right border
of the window tends to $+\infty$.

In this paper we have mostly studied the NLSE with an external potential
that has the form of a linear ramp with variable hight and slope. We have
applied an optimized fourth-order symplectic integrator in our simulations.
The integrator is optimized in the sense that it takes into account the
symplectic structure of the NLSE and the fact that the kinetic energy is
bilinear in $\Psi_x$. It has a very good long-time stability.

It is very easy to apply this algorithm to NLSEs with other external
potentials, such as one or more impurities, a series of step functions or
any smooth function. In the same way other Hamiltonian disturbances can be
treated, e.g.~terms like $|\Psi|^{2p}\Psi$. In order to study the collision
of two solitons (a fast one and a slow one) or of three solitons (a fast
one crossing two interacting slow ones) \cite{kivsh:86,frauenkron:96:2},
an integrator with good long-time stability is needed. We believe that
the integrator used in the present work can be very useful in further
simulations of partial differential equations of such a type. This
numerical method can also be easily and staightforward applied to the NLSE
with a varying nonlinear term or to higher dimensional and coupled NLSEs.
Thus the vector NLSE, which describes e.g.\ the propagation of light in a
birefringent optical fiber, can be simulated with such an algorithm. Also
soliton switching and splitting can be simulated by using this algorithm.

\section*{Acknowledgements}
The author would like to thanks Prof.\ P.\ Grassberger, Dr.\ Yu.S.\ 
Kivshar and B.A.\ Malomed for fruitful discussions.

This work was supported by DFG within the Graduiertenkolleg
``Feld\-theo\-re\-tische und numerische Methoden in der Elementarteilchen-
und Statistischen Physik", and within SFB 237.

\section*{Appendix}
\begin{appendix}
\section{Symplectic integration}

Here we briefly describe the integration method which we have used to
simulate the NLSE. 

As we have already mentioned, the NLSE is a Hamiltonian system. Thus, it is
natural to apply to it integration routines which were developed during the
recent years and whose main characteristic is that they preserve the
Hamiltonian structure \cite{yoshida:90,yoshida:93,serna:92}. The latter is
not true e.g.\ for standard methods as e.g.\ Runge-Kutta or
predictor-corrector.  Such `symplectic' integrators (the simplest of which
is the well known Verlet or `leap frog' algorithm) have been applied
already to the linear \cite{bandrauk:91,takahashi,frauenkron:94,rouhi} and
nonlinear
\cite{frauenkron:95,weideman:86,pathria:90,bandrauk:94,mclachlan:94}
Schr\"odinger equations.

The most popular algorithms of this type are split-operator methods.  They
depend on the Hamiltonian being a sum of two terms $A$ and $B$, each of
which can be integrated explicitly. Then one uses the
Baker-Campbell-Hausdorff theorem to approximate $e^{i(A+B)t}$ by a product
of factors $e^{i\alpha_k At}$ and $e^{i\beta_k Bt}$, where $\alpha_k$ and
$\beta_k$ are real numbers satisfying among others $\sum_k \alpha_k =
\sum_k \beta_k =1$. The error is given by higher order commutators of
$A$ and $B$. In particular, we apply a fourth-order method introduced
by McLachlan and Atela \cite{mclachlan:92} which is applicable if one of
the third order commutators vanished identically. We found that this
method should be applicable to our problem, and that it is indeed
numerically very precise, indicating that the McLachlan-Atela method is the
method of choice for a wide class of problems.

The NLSE is a classical Hamiltonian system with Poisson bracket
\begin{equation}
   \{\Psi^\ast(x),\Psi(y)\} = i\delta(x-y) 
\end{equation}
and Hamiltonian $H = E$. This implies in particular that it can be written
as
\begin{equation} 
   \dot{\Psi} = \{\Psi,H\} = \mathcal{H} \Psi \, ,
\end{equation}
where the linear (`Liouville') operator $\mathcal{H}$ is defined as $\mathcal{H}\;\cdot = \{\;\cdot\; , H\}$. Split-operator methods can be applied by
splitting $\mathcal{H} = \mathcal{T}+\mathcal{V}$, where $\mathcal{T}$ and $\mathcal{V}$
are the Liouvilleans corresponding to ${1\over 2} \int \dx |\partial_x
\Psi|^2$ and $\int \dx(- {1\over 2}|\Psi|^4+V|\Psi|^2)$,
\begin{equation}\label{TV}
   \mathcal{T}\Psi = {i\over 2} \partial_x^2 \Psi\;,\quad \mathcal{V}\Psi =
   i(|\Psi|^2\Psi-V\Psi) \;.
\end{equation}

In \cite{mclachlan:92}, a fourth-order algorithm was introduced which
minimizes the neglected fifth order terms in the Baker-Campbell-Hausdorff
formula for Hamiltonians for which
\begin{equation}\label{comm3}
    [[[\mathcal{T}, \mathcal{V}],\mathcal{V}],\mathcal{V}] \equiv 0\;.
\end{equation} 
This applies obviously to Hamiltonians with $T = {1\over 2}(p,M^{-1}p)$,
$V=V(x)$, with $M$ a constant mass matrix and $\{q_i,p_k\}=\delta_{ik}$,
since in the Baker-Campbell-Hausdorff formula each commutator with $V$ acts
as a derivative operator on any function of $p$. In \cite{frauenkron:94} it
was shown that this algorithm can also be applied to the linear
Schr\"odinger equation where it gives better performance than the general
fourth-order algorithm \cite{yoshida:90} which does not take into account
this special structure. In \cite{frauenkron:95} it was shown that
eq.~(\ref{comm3}) holds also for $\mathcal{T}$ and $\mathcal{V}$ defined in
eq.~(\ref{TV}) and the McLachlan-Atela method can thus also be used for the
NLSE.

The coefficients $\alpha_k$ and $\beta_k$ for the McLachlan-Atela method
are listed in \cite{mclachlan:92,frauenkron:94}. Our implementation
involves a spatial grid with Fourier transformation after each half step
\cite{frauenkron:94}.\footnote{As an alternative to the Fourier transform
  one can perform these steps also in $x$-space \cite{torcini:95}. Instead
  of a FFT, one has to perform a convolution with the Greensfunction of
  $\mathcal{T}$ in each time step. Practically, both methods are roughly
  equally fast.}

Since $\mathcal{T}$ and $\mathcal{V}$ both conserve the mass exactly, $N$ should
be conserved up to round-off errors. This was checked numerically, relative
errors typically were of order $10^{-11}$. Energy is not conserved exactly,
and its error was $\approx 10^{-7}$ after an evolution time $t=500$ with an
integration step $\Delta t=0.0025$. The precise value of the error depends
of course on the parameters of the soliton and on $x_0$. We checked
carefully that our final results were independent of the time step and of
the spatial discretization $\Delta x$. It was checked that the algorithm is
indeed fourth-order, and is more precise than the general fourth-order
symplectic \cite{yoshida:90} and the leap-frog (second-order symplectic)
algorithms.

The derivatives of $\Psi$ and $\Psi^\ast$, which have to be calculated
numerically, were computed in Fourier space, as this is much more precise
than taking finite differences in $x$-space.  Using the latter, the
relative error of the energy conservation would have been of the order
$10^{-4}$ instead of $10^{-7}$.

\end{appendix}

\newcommand{\noopsort}[1]{} \newcommand{\printfirst}[2]{#1}
  \newcommand{\singleletter}[1]{#1} \newcommand{\switchargs}[2]{#2#1}

\end{document}